\documentclass[11pt]{article}
\usepackage[cp1250]{inputenc}
\usepackage{amsmath} 
\usepackage{amssymb} 
\usepackage{a4wide} 
\usepackage[left=3cm]{geometry} 
\usepackage{braket}
\usepackage{slashed}
\usepackage{bbm}
\usepackage{comment}
\usepackage{yfonts}
\usepackage{simplewick}
\usepackage[boxsize=0.7em,centertableaux]{ytableau}
\usepackage{empheq}
\usepackage{tikz}
\usetikzlibrary{matrix,arrows}
\usetikzlibrary{decorations.markings, decorations.pathreplacing}

\numberwithin{equation}{section}

\DeclareMathOperator{\res}{res}

\DeclareMathOperator{\Sym}{Sym}

\title{Instanton R-matrix and W-symmetry}
\author{Tomáš Procházka}

\begin{document}

\bibliographystyle{hieeetr}

\vskip 2.1cm

\centerline{\large \bf Instanton $R$-matrix and $\mathcal{W}$-symmetry}
\vspace*{8.0ex}

\centerline{\large \rm Tom\'{a}\v{s} Proch\'{a}zka\footnote{Email: {\tt tomas.prochazka@lmu.de}}}

\vspace*{8.0ex}

\centerline{\large \it Arnold Sommerfeld Center for Theoretical Physics}
\centerline{\large \it Ludwig Maximilian University of Munich}
\centerline{\large \it Theresienstr. 37, D-80333 München, Germany}
\vspace*{2.0ex}

\vspace*{6.0ex}

\centerline{\bf Abstract}
\bigskip

We study the relation between $\mathcal{W}_{1+\infty}$ algebra and Arbesfeld-Schiffmann-Tsymbaliuk Yangian using the Maulik-Okounkov R-matrix. The central object linking these two pictures is the Miura transformation. Using the results of Nazarov and Sklyanin we find an explicit formula for the mixed R-matrix acting on two Fock spaces associated to two different asymptotic directions of the affine Yangian. Using the free field representation we propose an explicit identification of Arbesfeld-Schiffmann-Tsymbaliuk generators with the generators of Maulik-Okounkov Yangian. In the last part we use the Miura transformation to give a conformal field theoretic construction of conserved quantities and ladder operators in the quantum mechanical rational and trigonometric Calogero-Sutherland models on which a vector representation of the Yangian acts.

 \vfill \eject

\tableofcontents

\setcounter{footnote}{0}

\newpage

\section{Introduction}
$\mathcal{W}$-algebras are remarkable algebraic structures introduced first by Zamolodchikov \cite{Zamolodchikov:1985wn}. The algebra that he was studying was an extension of the Virasoro algebra underlying two-dimensional conformal field theory by an additional generator of spin $3$. Since then there have been many applications of these algebras in various areas of mathematical physics. Among the oldest are applications to integrable hierarchies of partial differential equations \cite{DiFrancesco:1990qr}, matrix integrals \cite{Kharchev:1992iv}, topological strings \cite{Aganagic:2003qj} or in the quantum Hall effect \cite{Cappelli:1992yv,Cappelli:1995ts}.

More recently there are two directions of research where $\mathcal{W}$-algebras play a prominent role. The first one is the $\mathrm{AdS}_3/\mathrm{CFT}_2$ duality with higher spin symmetries \cite{Campoleoni:2010zq,Campoleoni:2011hg,Gaberdiel:2010pz,Gaberdiel:2011zw}. The cosmological Einstein gravity in three dimensions can be formulated as a Chern-Simons theory with the gauge group being two copies of $SL(2,\mathbbm{R})$ \cite{Achucarro:1987vz,Witten:1988hc}. Replacing the $SL(2,\mathbbm{R})$ gauge group by $SL(N,\mathbbm{R})$ extends the gravity theory to a theory of higher spins. The algebra of asymptotic symmetries which in the case of Einstein gravity is the Virasoro algebra \cite{Brown:1986nw} is in this case extended to a $\mathcal{W}_N$ algebra. These asymptotic symmetry algebras are interpreted as symmetry algebras of holographic dual two-dimensional conformal field theories. There have been many extensions of this class of holographic dualities for various other gauge groups \cite{Gaberdiel:2011nt,Candu:2012ne} or including supersymmetry \cite{Candu:2012jq,Candu:2012tr,Gaberdiel:2013vva,Beccaria:2014jra,Gaberdiel:2014yla}. Later this program was extended to stringy holography \cite{Gaberdiel:2014cha,Gaberdiel:2015mra,Gaberdiel:2015wpo,Eberhardt:2017pty,Eberhardt:2018plx,Eberhardt:2018ouy}. In this series of papers the authors found an explicit holographic duality involving string theory on $AdS_3 \times S_3 \times T^4$ at special point in the moduli space where both sides of the duality have tractable description.

Another area where $\mathcal{W}$-algebras show up are 4-dimensional $\mathcal{N}=2$ supersymmetric field theories in connection with their BPS states. The AGT correspondence \cite{Alday:2009aq,Wyllard:2009hg} relates the instanton partition functions of 4d supersymmetric gauge theories \cite{Nekrasov:2002qd,Nekrasov:2003rj} with $SU(N)$ gauge group to two-dimensional conformal blocks with $\mathcal{W}_N$ symmetry. Geometrically $\mathcal{W}$-algebras act on equivariant cohomology of instanton moduli spaces \cite{schifvas,Maulik:2012wi,Braverman:2014xca,Tachikawa:2014dja,Rapcak:2018nsl}. $\mathcal{W}$-algebras can be also seen as a subsector of local fields of superconformal field theories in 4 and 6 dimensions \cite{Beem:2013sza,Beem:2014kka} whose traces can in turn be seen also holographically \cite{Bastianelli:1999en,Corrado:1999pi} in compactifications of 11-dimensional supergravity. The index calculations in \cite{Cecotti:2010fi,Cordova:2015nma} link $\mathcal{W}$-algebra characters to wall-crossing formulas \cite{Kontsevich:2008fj,Kontsevich:2010px}. Last but not least, $\mathcal{W}$-algebras do not only act on equivariant cohomology of instanton moduli spaces but also on moduli spaces of Higgs bundles \cite{Fredrickson:2017jcf} which again are tightly connected with the $\mathcal{N}=2$ quantum field theories \cite{Gaiotto:2009hg}.

There is an interesting two-parametric family of $\mathcal{W}$-algebras $\mathcal{W}_{1+\infty}$ which is generated by fields of dimension $1,2,3,\ldots$ \cite{Hornfeck:1994is,Gaberdiel:2012ku,Linshaw:2017tvv}. For special values of parameters its quotients are $\mathcal{W}_{N}$ algebras associated to $A_k$ series of simple Lie algebras. It also contains two-parametric family of even spin $\mathcal{W}_{ev \infty}$ whose quotients are orthosymplectic $\mathcal{W}$-algebras associated $B_k$, $C_k$ and $D_k$ series \cite{Candu:2012ne,Kanade:2018qut}. The $\mathcal{W}_{1+\infty}$ family admits a completely different description as Yangian of $\widehat{\mathfrak{gl}(1)}$, an associative algebra given by generators and relations \cite{schifvas:2011,arbesschif,tsymbaliuk2017affine}. The map between the two pictures is non-local from VOA point of view and manifests integrability on the Yangian side. Although the map is known explicitly at the level of generators \cite{Prochazka:2015deb}, there is a more conceptual understanding of the transformation using the Maulik-Okounkov instanton $\mathcal{R}$-matrix \cite{Maulik:2012wi,Smirnov:2013hh}. The aim of this article is to discuss this explicitly.

The central object linking the Yangian and $\mathcal{W}$-algebra is the Miura transformation ($GL(n)$-oper) \cite{Fateev:1987zh,luk1988quantization,Lukyanov:1990tf,Bouwknegt:1992wg}. From $\mathcal{W}$-algebra point of view it provides a free field representation of the algebra. The Miura transformation shows that one can think of $\mathcal{W}_N$-algebra as being a quantization of the space of $N$-th order differential operators \footnote{An analytic continuation of this to pseudo-differential operators and $\mathcal{W}_{\infty}$ is considered in \cite{Khesin:1994ey, Linshaw:2019xaq}.}. The specific free field embedding of $\mathcal{W}_N$ in the Fock space of $N$ free bosons depends on the ordering of the free fields. The observation of Maulik and Okounkov is that the Fock space operator intertwining one embedding with the another one satisfies the Yang-Baxter equation \cite{Maulik:2012wi}. Knowing this, we can apply the machinery of algebraic Bethe ansatz \cite{Faddeev:1987ih,Faddeev:1996iy,Nepomechie:1998jf} to study this algebra. In particular, the generators and relations given by Arbesfeld, Schiffmann and Tsymbaliuk should follow from the Yang-Baxter equation with Maulik-Okounkov instanton $\mathcal{R}$-matrix.

Currently the expressions for the instanton $\mathcal{R}$-matrix in the bosonic case are known explicitly to first few orders in the large spectral parameter expansion \cite{Zhu:2015nha}. There is also a fermionic expression for $\mathcal{R}$-matrix given in \cite{Smirnov:2013hh} which is however rather complicated. The aim of this work is to derive another expression for $\mathcal{R}$-matrix, study the resulting relations in the Yangian algebra and compare it to Arbesfeld-Schiffmann-Tsymbaliuk presentation. Actually, what we find is a formula for $\mathcal{R}$-matrix where each of the representation spaces is a Fock space, but these representations are inequivalent representations of $\mathcal{W}_{1+\infty}$. In this case, we can use the expressions for higher Jack Hamiltonians found by Nazarov and Sklyanin \cite{Nazarov:2012gn,nazarovskl} to write a closed-form formula for the $\mathcal{R}$-matrix.

\subsection*{Overview}
Let us now summarize in more detail the content of this article.

In section \ref{secfreefield} we review the free field representations of $\mathcal{W}_{1+\infty}$. In particular, there are three inequivalent representations in Fock space of a single boson associated to three parameters exchanged by the triality symmetry. The corresponding Miura factors were introduced already in \cite{Prochazka:2018tlo}, but here we rewrite them in compact form (\ref{miuracompact}) as suggested by A. Litvinov. We study in detail conformal transformation properties of Miura operators. Written in terms of a differential operator, the resulting expression (\ref{totalmiura3}) agrees with the classical case considered in \cite{DiFrancesco:1990qr}. The expression (\ref{miuraconftransf}) generalizes it to the case of pseudo-differential operator where the elementary Miura factors are of different type.

The next section introduce the $\mathcal{R}$-matrix following \cite{Maulik:2012wi, Zhu:2015nha}. We study the $\mathcal{R}$-matrix not only in the case where both of the representation spaces on which $\mathcal{R}$ acts are the same representation, but also $\mathcal{R}$-matrices of a mixed type where both spaces are still Fock spaces but associated to different asymptotic directions in $\mathcal{W}_{1+\infty}$ parameter space. Next, we show how to evaluate matrix elements of Fock-space $\mathcal{R}$-matrix from its definition, without using any expansion in large spectral parameter. Finally, following the logic of algebraic Bethe ansatz, we consider three special matrix elements of $\mathcal{R}$-matrix acting between auxiliary Fock space and an arbitrary quantum space. These matrix elements will play the role of Arbesfeld-Schiffmann-Tsymbaliuk Hamiltonians and raising and lowering operators. Using the Yang-Baxter equation we derive some relations satisfied by these operators.

Next, in section \ref{secboson} we consider a special case where the quantum space is a single free boson. We find that the generating function of Hamiltonians is diagonalized by Jack polynomials. Interesting Hamiltonians acting on Jack polynomials were studied by Nazarov and Sklyanin \cite{Nazarov:2012gn,nazarovskl} and we use their expressions for Hamiltonians to reconstruct the full mixed $\mathcal{R}$-matrix (\ref{rmix}). We next turn to ladder operators and comparing them to ladder operators studied in \cite{Prochazka:2015deb} we find a candidate for a map between these (\ref{EHtoe1},\ref{EHtoe2},\ref{FHtof}). We conclude the section by mentioning a nice determinantal formula for another set of commuting Hamiltonians found by Nazarov and Sklyanin. In combination with the fermionic expression for $\mathcal{R}$-matrix these give a quantum analogue of Szeg\"{o} formula.

In section \ref{secfermion} we derive another formula for the mixed $\mathcal{R}$-matrix. We start by fermionic representation of the commuting Hamiltonians for a special choice of parameters of $\mathcal{W}_{1+\infty}$. To find a formula for any value of $\mathcal{W}_{1+\infty}$ parameters, we use the result of Nazarov-Sklyanin that in the bosonic picture the deformation of parameters can be achieved purely by rescaling the normalization of Fock oscillators once the operators are written in a normal ordered form. On the other hand, after we bosonize the fermionic fields, the resulting vertex operators can be brought to a normal ordered form in a well-known way by Wick theorem. In section \ref{secastrelations} we test some of the relations between the Yangian generators and the corresponding relations of Arbesfeld-Schiffmann-Tsymbaliuk.

In the final section we interpret the elementary Miura factor as being a transfer matrix in Fock representation and in representation by differential operators acting on CFT worldsheet. In the case of a cylinder, we find Hamiltonians of Calogero-Sutherland model together with ladder operators satisfying the Yangian commutation relations. Geometrically, we consider  correlation functions with $n$ insertions of Miura operators on cylinder and a special in and out state at plus and minus infinity. Each choice of in and out states corresponds to one matrix element of the $\mathcal{R}$-matrix which is a differential operator acting on the space of positions of Miura insertions (i.e. the moduli space of a cylinder with $n$ punctures). The Yangian algebra encodes the Ward identities for this class of correlation functions as we vary the in and out states. Choosing just a one insertion leads to a vector representation of $\mathcal{W}_{1+\infty}$ which is the simplest known representation of the algebra \cite{tsymbaliuk2017affine}. Although we have not studied what happens for higher genus surfaces (where one would need to understand how the handle insertion interacts with the $\mathcal{R}$-matrix), it is nice to see that $n$-point functions of the elementary Miura factor on torus reproduce the Hamiltonians of elliptic Calogero-Moser systems \cite{Polychronakos:2018gfz}.

\section{Miura transformation}
\label{secfreefield}

One of the possible ways of defining $\mathcal{W}_N$ algebras is starting from their free field representation \cite{Fateev:1987zh,	luk1988quantization, Prochazka:2014gqa}. We first consider $N$ free $\hat{\mathfrak{u}}(1)$ currents $J_j(z)$ with OPE
\begin{equation}
J_j(z) J_k(w) \sim \frac{\delta_{jk}}{(z-w)^2}
\end{equation}
and define an operator
\begin{equation}
\label{basicmiura}
\mathcal{L}(z) = (\alpha_0 \partial + J_1(z)) (\alpha_0 \partial + J_2(z)) \cdots (\alpha_0 \partial + J_N(z)) \equiv \sum_{k=0}^N U_k(z) (\alpha_0 \partial)^{N-k}.
\end{equation}
This operator will play a central role in the following. Here $\alpha_0$ is a free parameter which will be later related to the central charge of the algebra. The non-trivial fact is that expressing the local fields $U_j(z)$ in terms of the free currents $J_k(z)$ as in (\ref{basicmiura}), the fields $U_j$ themselves generate a closed algebra under operator product expansions. This algebra is by definition $\widehat{\mathfrak{u}}(1) \times \mathcal{W}_N \equiv \mathcal{Y}_{0,0,N}$. The currents $U_j(z)$ do not transform as primary fields under conformal transformations, but perhaps surprisingly their operator product expansions have purely quadratic non-linearity \cite{luk1988quantization, Prochazka:2014gqa}. This is one of signs of the connection to integrability, where the algebras with quadratic non-linearity appear naturally.

We can split the $N$ free bosons in (\ref{basicmiura}) into two groups of $N_1$ and $N_2$ bosons. Multiplying two Miura operators associated to $N_1$ and $N_2$ bosons and passing the derivatives to the right, we find a coproduct in $\mathcal{W}_{1+\infty}$ which embeds \cite{Prochazka:2014gqa}
\begin{equation}
\mathcal{Y}_{0,0,N_1+N_2} \subset \mathcal{Y}_{0,0,N_1} \times \mathcal{Y}_{0,0,N_2}.
\end{equation}
This fusion operation fixes the value of $\alpha_0$ parameter, i.e. the ratios of $\lambda_j$ parameters introduced later stay the same and the vector $(\lambda_1,\lambda_2,\lambda_3)$ is thus additive under the fusion \cite{Prochazka:2014gqa}.

Another important point to mention is that the way $\mathcal{W}_N$ is embedded in the bosonic Fock space depends on the ordering of free fields in (\ref{basicmiura}). It was noticed in \cite{Maulik:2012wi, Zhu:2015nha} that the intertwining operator between different embeddings satisfies Yang-Baxter equation. The aim of this work is to study this $\mathcal{R}$-matrix and use it to connect $\mathcal{W}$-algebras to their Yangian description \cite{schifvas, arbesschif, tsymbaliuk2017affine, Prochazka:2015deb}.

\subsection{Other triality frames}

In \cite{Hornfeck:1994is,Gaberdiel:2012ku,Linshaw:2017tvv} a two-parametric family of algebras called $\mathcal{W}_{\infty}$ was studied which interpolates between all the $\mathcal{W}_N$ algebras. Unlike the linear versions of $\mathcal{W}_{\infty}$ constructed in \cite{Pope:1989sr, Pope:1990kc}, this two-parametric family has all $\mathcal{W}_N$ algebras with an arbitrary value of the central charge as its truncations. It is generated by fields of spin $2, 3, \ldots$ (with one generator of every spin). A very surprising property found by \cite{Gaberdiel:2012ku} was the triality symmetry of the algebra: parametrizing $\mathcal{W}_{\infty}$ in terms of the central charge $c$ and the rank-like parameter $\lambda$, for each value of $c$ there are generically three values of $\lambda$ which give the same structure constants. This has important consequences for the representation theory of the algebra and points towards to the connection to topological strings and affine Yangian picture \cite{Prochazka:2014gqa, Prochazka:2015deb}.

To make the triality symmetry manifest, it is useful to parametrize the algebra in terms of three values $\lambda_j$. They are related by the equation
\begin{equation}
\label{lambdaconstraint}
\frac{1}{\lambda_1} + \frac{1}{\lambda_2} + \frac{1}{\lambda_3} = 0.
\end{equation}
Furthermore, the central charge of the stress-energy tensor of $\mathcal{W}_{\infty}$ is parametrized as \cite{Prochazka:2014gqa}
\begin{equation}
c = (\lambda_1-1)(\lambda_2-1)(\lambda_3-1).
\end{equation}
The relation between these and the parameters $\alpha_0$ and $N$ appearing in (\ref{basicmiura}) is
\begin{equation}
\lambda_3 = N
\end{equation}
and
\begin{equation}
c = (N-1)(1-N(N+1)\alpha_0^2).
\end{equation}
Note that this choice of the identification of parameters manifestly breaks the triality symmetry, i.e. the Miura transformation (\ref{basicmiura}) is written with the choice of 3rd direction as the preferred one. From the triality symmetry we expect that there should be also free field representations corresponding to integer values of $\lambda_1$ or $\lambda_2$. That this is indeed the case was verified in \cite{Prochazka:2018tlo}.

The truncations of $\mathcal{W}_{1+\infty}$ to $\mathcal{Y}_{0,0,N} \equiv \hat{\mathfrak{u}}(1) \times \mathcal{W}_N$ are not the only possible truncations of the algebra. By studying singular vectors of the vacuum representation \cite{Prochazka:2014gqa} and independently from the gauge theory construction \cite{Gaiotto:2017euk, Prochazka:2017qum} it was understood that we have a family of truncations $\mathcal{Y}_{N_1,N_2,N_3}$ parametrized by three non-negative integers $N_1, N_2$ and $N_3$. If the $\lambda$-parameters of $\mathcal{W}_{1+\infty}$ satisfy the constraint
\begin{equation}
\frac{N_1}{\lambda_1} + \frac{N_2}{\lambda_2} + \frac{N_3}{\lambda_3} = 1,
\end{equation}
there appears a singular vector in the vacuum representation at level $(N_1+1)(N_2+1)(N_3+1)$ and the whole infinitely-generated $\mathcal{W}_{1+\infty}$ can be truncated to a subalgebra generated by fields of spin $1, 2, \ldots, (N_1+1)(N_2+1)(N_3+1)-1$ \footnote{If one of $N_j$ is zero, this is the lowest level singular vector that appears for generic values of the central charge. If all $N_j$ are positive, due to (\ref{lambdaconstraint}) there is a singular vector at lower level in the vacuum module. If two of $N_j$ parameters vanish, the algebra is freely generated.}. There is a nice combinatorial description of the truncation in terms of plane partitions (box-counting) which is discussed in \cite{Gaiotto:2017euk, Prochazka:2017qum}.

The free field representation of $\mathcal{Y}_{N_1,N_2,N_3}$ was constructed in \cite{Prochazka:2018tlo}. The idea is to consider a simple modification of (\ref{basicmiura}) where we generalize the elementary factor
\begin{equation}
\label{elementarymiura}
\mathcal{L}^{(3)}(z) \equiv \alpha_0 \partial + J^{(3)}(z) = : e^{-\frac{1}{h_3} i \phi^{(3)}(z)} (\alpha_0 \partial) e^{\frac{1}{h_3} i\phi^{(3)}(z)} :
\end{equation}
associated to the third direction and to $\mathcal{Y}_{0,0,1}$ algebra by finding an analogous factors $\mathcal{L}^{(1)}$ and $\mathcal{L}^{(2)}$. The only tricky point is that the quadratic basis of $U_j$ fields introduced in (\ref{basicmiura}) are also associated to the 3rd direction so if we try to find an expression for $\mathcal{Y}_{1,0,0}$ or $\mathcal{Y}_{0,1,0}$ generators in terms of $U_j$ fields, there will be an infinite number of non-trivial $U_j$ fields. Related to this, the leading power (order) of the differential operator $\mathcal{L}$ in this basis is $\lambda_3$ which does not take a positive integral value for $\mathcal{Y}_{1,0,0}$ or $\mathcal{Y}_{0,1,0}$ so instead of a differential operator we have to consider a formal pseudo-differential operator. With this in mind, we can write \cite{Prochazka:2018tlo}
\begin{equation}
\label{generalbasicmiura}
\mathcal{L}^{(\tau)}(z) \equiv (\alpha_0 \partial)^{\frac{h_{\tau}}{h_3}} + \sum_{k=1}^{\infty} U_k^{(\tau)}(z) (\alpha_0 \partial)^{\frac{h_{\tau}}{h_3}-k}
\end{equation}
with
\begin{equation}
U_j^{(\tau)} = \prod_{k=1}^{j-1} \left(1-\frac{k h_3}{h_\tau}\right) \sum_{m_1 + 2m_2 + \ldots + jm_j = j} \prod_{k=1}^j \frac{1}{m_k! k^{m_k}} \left(\frac{h_{\tau}^{k-1}}{(k-1)!} \partial^{k-1} J^{(\tau)} \right)^{m_k}.
\end{equation}
By an observation of A. Litvinov, this can be also written in terms of free boson normal ordering as
\begin{equation}
\label{miuracompact}
\mathcal{L}^{(\tau)}(z) = : e^{-\frac{1}{h_\tau} i\phi^{(\tau)}(z)} (\alpha_0 \partial)^{\frac{h_\tau}{h_3}} e^{\frac{1}{h_\tau} i\phi^{(\tau)}(z)} :
\end{equation}
i.e. as pseudo-differential operator $\partial^{\frac{h_{\tau}}{h_3}}$ dressed by a free boson vertex operator. The free bosons are normalized as
\begin{equation}
\label{freefieldnorm}
J_j^{(\tau_j)}(z) J_k^{(\tau_k)}(w) \sim -\frac{h_{\tau_j}}{h_1 h_2 h_3} \frac{\delta_{jk}}{(z-w)^2}.
\end{equation}
with $J(z) = i\partial\phi(z)$ and the Yangian parameters are introduced via \cite{Prochazka:2015deb}
\begin{equation}
\lambda_j = -\frac{\psi_0 h_1 h_2 h_3}{h_j}.
\end{equation}
Having found these three basic free field representations, a free field representation of an arbitrary $\mathcal{Y}_{N_1,N_2,N_3}$ can be obtained by taking $N_1+N_2+N_3$ free bosons normalized as in (\ref{freefieldnorm}) and multiplying the corresponding simple Miura factors (\ref{generalbasicmiura}).

As already mentioned, the way $\mathcal{Y}_{N_1,N_2,N_3}$ is embedded in the bosonic Fock space depends on the choice of ordering of basic Miura factors. Since these different choices of the order are equivalent, there should be a Fock space operator that intertwines between these embedding. For an elementary permutation of two neighboring Miura factors this will be the $\mathcal{R}$-matrix that is the main subject of this article.

\subsection{Conformal transformations}
\label{secconftrans}

In the following, it will be useful to understand how the fields appearing in the Miura transformation transform under conformal transformations. Let's first focus on the case of $\mathcal{Y}_{0,0,N}$ algebras with Miura transformation (\ref{basicmiura}). The algebra $\mathcal{W}_{1+\infty}$ has a unique stress-energy tensor with respect to which the spin $1$ current is a primary of dimension $1$. The formula for this stress-energy tensor is
\begin{equation}
T(z) \equiv T_{1+\infty}(z) = -U_2(z) + \frac{(N-1)\alpha_0}{2} \partial U_1(z) + \frac{1}{2} (U_1 U_1)(z).
\end{equation}

\paragraph{Stress-energy tensor as generator of conformal transformations}
One of the important roles played by $T(z)$ is that it is a generator of the conformal transformations. Consider an infinitesimal conformal transformation
\begin{equation}
z \to \tilde{z} = z + \epsilon(z) + \mathcal{O}(\epsilon^2).
\end{equation}
Under this transformation, the fields transform such that
\begin{equation}
\tilde{\phi}(\tilde{z}) - \phi(\tilde{z}) = -\oint_w \frac{dz}{2\pi i} \epsilon(z) T(z) \phi(w) + \mathcal{O}(\epsilon^2).
\end{equation}
As an example, consider a primary field $\phi(z)$ of dimension $h$. Being primary of weight $h$ means that under conformal transformations it transforms as
\begin{equation}
\phi(z) \to \tilde{\phi}(\tilde{z}) = \left( \frac{d\tilde{z}}{dz} \right)^{-h} \phi(z).
\end{equation}
For an infinitesimal transformation, we find
\begin{equation}
\tilde{\phi}(\tilde{z}) - \phi(\tilde{z}) = -h\epsilon^\prime(z)\phi(z) - \epsilon(z)\phi^\prime(z) + \ldots
\end{equation}
which is indeed equal to
\begin{equation}
-\oint_w \frac{dz}{2\pi i} \epsilon(z) T(z) \phi(w)
\end{equation}
if we use the OPE of the stress-energy tensor primary field $\phi(w)$
\begin{equation}
T(z) \phi(w) \sim \frac{h \phi(w)}{(z-w)^2} + \frac{\partial \phi(w)}{z-w} + reg.
\end{equation}

\paragraph{More complicated transformation properties}
Not all the fields transform as simply as the primary fields. Let us consider two examples of fields that transform in more complicated way. First is the most well-known transformation property of the stress-energy tensor itself. Due to OPE
\begin{equation}
\label{stresstensorope}
T(z) T(w) \sim \frac{c/2}{(z-w)^4} + \frac{2T(w)}{(z-w)^2} + \frac{\partial T(w)}{z-w}
\end{equation}
which has an anomalous quartic term proportional to the central charge, the transformation of $T(z)$ is
\begin{equation}
T(z) \to \tilde{T}(\tilde{z}) = \left(\frac{d\tilde{z}}{dz}\right)^{-2} T(z) - \frac{c}{12} \left(\frac{d\tilde{z}}{dz}\right)^{-2} \left[ \left(\frac{d^3 \tilde{z}}{dz^3}\right) \left( \frac{d\tilde{z}}{dz} \right)^{-1} - \frac{3}{2} \left( \frac{d^2 \tilde{z}}{dz^2} \right)^2 \left( \frac{d\tilde{z}}{dz} \right)^{-2} \right].
\end{equation}
The second term on the right-hand side is the anomalous term coming from the quartic pole of OPE of $T(z)$ with itself. It is proportional to the Schwarzian derivative of the function $\tilde{z}(z)$.

Another example is the $\hat{\mathfrak{u}}(1)$ current $J(z)$ with the OPE
\begin{equation}
T(z) J(w) \sim \frac{2\alpha}{(z-w)^3} + \frac{J(w)}{(z-w)^2} + \frac{\partial J(w)}{z-w}.
\end{equation}
The linear and quadratic poles are just those of ordinary spin $1$ primary, but the cubic pole is responsible for the anomalous transformation property
\begin{equation}
\label{anomalousspin1}
J(z) \to \tilde{J}(\tilde{z}) = \left( \frac{d\tilde{z}}{dz} \right)^{-1} J(z) - \alpha \left( \frac{d\tilde{z}}{dz} \right)^{-2} \left( \frac{d^2\tilde{z}}{dz^2} \right).
\end{equation}
In a conformal field theory with a spin $1$ primary current $J(z)$ we can deform the stress-energy tensor by redefining $T \to T - \alpha \partial J$. The new Feigin-Fuchs stress-energy tensor still satisfies the OPE (\ref{stresstensorope}) (with modified value of the central charge) but the field $J(z)$ is no longer primary and has exactly the anomalous transformation (\ref{anomalousspin1}).

\paragraph{Transformation of $U_j$ fields}
Let us now return to the Miura transformation (\ref{basicmiura}). As already mentioned, the fields $U_j$ do not transform as primary fields under the conformal transformations. The stress-energy tensor $T(z) \equiv T_{1+\infty}(z)$ takes the form
\begin{equation}
T_{1+\infty}(z) = -U_2(z) + \frac{(N-1)\alpha_0}{2} \partial U_1(z) + \frac{1}{2} (U_1 U_1)(z).
\end{equation}
The operator product expansions of $T_{1+\infty}$ with $U_j$ has non-trivial poles of order up to $j+2$, but their coefficients are proportional to $U_k(w)$ with an exception of the linear pole which equals $\partial U_j$. This means that the conformal transformations of $U_j$ fields form a triangular matrix which mixes $U_j$ fields with $U_k$, $k < j$. As an illustration, for dimension $2$ field $U_2$ we find
\begin{align}
\widetilde{U_2}(\tilde{z}) = & \left( \frac{\partial \tilde{z}}{\partial z} \right)^{-2} U_2(z) - \frac{(N-1)\alpha_0}{2} \left( \frac{\partial^2 \tilde{z}}{\partial z^2} \right) \left( \frac{\partial \tilde{z}}{\partial z} \right)^{-3} U_1(z) \\
\nonumber
& - \frac{(N-1)N(N+1)\alpha_0^2}{12} \left( \frac{\partial \tilde{z}}{\partial z} \right)^{-2} \left[ \left(\frac{d^3 \tilde{z}}{dz^3}\right) \left( \frac{d\tilde{z}}{dz} \right)^{-1} - \frac{3}{2} \left( \frac{d^2 \tilde{z}}{dz^2} \right)^2 \left( \frac{d\tilde{z}}{dz} \right)^{-2} \right].
\end{align}
One can deduce this from the conformal transformation property of $U_1$ which is primary of spin $1$ and of the stress-energy tensor itself. An alternative way is to verify that this expression behaves well under the composition of conformal transformations and that infinitesimally it agrees with the OPE with $T_{1+\infty}$.

We could proceed analogously for higher $U_j$ fields, but it is easier to find first how the free boson fields transform under the conformal transformations. In terms of $J_j$ fields of (\ref{basicmiura}), the stress-energy tensor is
\begin{equation}
T_{1+\infty}(z) = \frac{1}{2} \sum_{j=1}^N (J_j J_j)(z) + \frac{\alpha_0}{2} \sum_{j=1}^N (N+1-2j) \partial J_j(z).
\end{equation}
The OPE with $J_j$ field is thus
\begin{equation}
T_{1+\infty}(z) J_j(w) \sim -\frac{\alpha_0(N+1-2j)}{(z-w)^3} + \frac{J_j(w)}{(z-w)^2} + \frac{\partial J_j(w)}{z-w}
\end{equation}
which means that under the conformal transformations $J_j(z)$ transform as
\begin{equation}
\tilde{J}_j(\tilde{z}) = \left(\frac{d\tilde{z}}{dz}\right)^{-1} J_j(z) + \frac{\alpha_0(N+1-2j)}{2} \left(\frac{d\tilde{z}}{dz}\right)^{-2} \left(\frac{d^2\tilde{z}}{dz^2}\right).
\end{equation}
Let us emphasize that this anomalous OPE depends on the ordering of $J_j$ fields in the definition of the Miura transformation (which is in contrast with $U_j$ fields whose transformation properties are governed by $\mathcal{W}_{1+\infty}$ algebra). This is because the expression for $T_{1+\infty}$ itself depends via Feigin-Fuchs background charge term on the way we ordered the free boson fields.

For our fixed ordering (where $J_1$ appears on the left), the elementary Miura factor transforms as
\begin{align}
\nonumber
\alpha_0 \partial_z + J_j(z) \to & \, \alpha_0 \partial_{\tilde{z}} + \tilde{J}_j(\tilde{z}) \\
\nonumber
= & \left( \frac{d\tilde{z}}{dz}\right)^{-1} \left[ \alpha_0 \partial_z + J_j(z) + \frac{\alpha_0(N+1-2j)}{2} \left( \frac{d\tilde{z}}{dz}\right)^{-1} \left( \frac{d^2\tilde{z}}{dz^2}\right) \right] \\
= & \left( \frac{d\tilde{z}}{dz}\right)^{-1-\frac{N+1-2j}{2}} \left[ \alpha_0 \partial_z + J_j(z)  \right] \left( \frac{d\tilde{z}}{dz}\right)^{\frac{N+1-2j}{2}}.
\end{align}
Multiplying these elementary factors, we see that the intermediate Jacobian factors cancel out and the total Miura $\mathcal{L}(z)$ has a nice transformation property
\begin{equation}
\label{totalmiura3}
\mathcal{L}(z) \to \tilde{\mathcal{L}}(\tilde{z}) = \left( \frac{d\tilde{z}}{dz} \right)^{-\frac{N+1}{2}} \mathcal{L}(z) \left( \frac{d\tilde{z}}{dz} \right)^{-\frac{N-1}{2}}.
\end{equation}
This agrees with \cite{DiFrancesco:1990qr} where the authors study the classical transformation properties of the $\mathcal{W}_N$ generators. What we see now is that these transformation properties at the level of $\mathcal{L}(z)$ are not modified in the quantum version of the algebra.

Another thing to be emphasized at this point is that the transformation of $\mathcal{L}(z)$ or $U_j(z)$ fields does not depend on the choice of ordering of the free fields. If we ordered the $J_j$ fields in a different way, the expression for $T_{1+\infty}$ in terms of $J_j$ would be modified in such a way that the full $\mathcal{L}(z)$ would still transform in the same way for all orderings. This is simply a consequence of the already mentioned fact that the transformation properties of $U_j$ fields which determine that of $\mathcal{L}(z)$ are independent of the particular free field representation.

\paragraph{Transformation of $\mathcal{L}$ for any type of bosons}
Let us now generalize the previous discussion to the case where the free bosons are not necessarily of the 3rd type, i.e. to the case of $\mathcal{Y}_{N_1,N_2,N_3}$. Consider the ordering
\begin{equation}
\mathcal{L}(z) = \mathcal{L}_1^{(\tau_1)}(z) \mathcal{L}_2^{(\tau_2)}(z) \dots \mathcal{L}_N^{(\tau_n)}(z)
\end{equation}
where $n \equiv N_1+N_2+N_3$. $N_1$ is the number of $j$ for which $\tau_j=1$ etc. The stress energy tensor $T_{1+\infty}(z)$ in terms of free fields $J_j^{(\tau_j)}(z)$ is given by \cite{Prochazka:2018tlo}
\begin{equation}
T_{1+\infty}(z) = -\frac{1}{2} \sum_{j=1}^n \frac{h_1 h_2 h_3}{h_{\tau_j}} (J_j J_j)(z) - \frac{1}{2} \sum_{j<k} h_{\tau_j} \partial J_k(z) + \frac{1}{2} \sum_{j>k} h_{\tau_j} \partial J_k(z)
\end{equation}
which reduces to the previous expression if all $\tau_j = 3$. The dependence on $h_j$ parameters is also consistent with the scaling symmetry of $\mathcal{W}_{1+\infty}$ discussed in \cite{Prochazka:2015deb}. OPE of $T_{1+\infty}$ with $J_j$ is easy to evaluate:
\begin{equation}
T_{1+\infty}(z) J_j(w) \sim \frac{h_{\tau_j} (\sum_{k>j} h_{\tau_k} - \sum_{k<j} h_{\tau_k})}{h_1 h_2 h_3} \frac{1}{(z-w)^3} + \frac{J_j(w)}{(z-w)^2} + \frac{\partial J_j(w)}{z-w}.
\end{equation}
This means that under the conformal transformations the currents $J_j$ transform as
\begin{equation}
J_j(z) \to \tilde{J_j}(\tilde{z}) = \left( \frac{d\tilde{z}}{dz} \right)^{-1} J_j(z) - \frac{h_{\tau_j} (\sum_{k>j} h_{\tau_k} - \sum_{k<j} h_{\tau_k})}{2h_1 h_2 h_3} \left( \frac{d\tilde{z}}{dz} \right)^{-2} \left( \frac{d^2\tilde{z}}{dz^2} \right).
\end{equation}
Using this, we can verify that the individual Miura factors transform as
\begin{equation}
\mathcal{L}^{(\tau_j)}_j(z) \to \tilde{\mathcal{L}}^{(\tau_j)}_j(\tilde{z}) = \left( \frac{d\tilde{z}}{dz}\right)^{-\frac{1}{2} + \frac{1}{2} \sum_{k<j} \frac{h_{\tau_k}}{h_3} - \frac{1}{2} \sum_{k \geq j} \frac{h_{\tau_k}}{h_3}} \mathcal{L}_j(z) \left( \frac{d\tilde{z}}{dz}\right)^{\frac{1}{2} - \frac{1}{2} \sum_{k \leq j} \frac{h_{\tau_k}}{h_3} + \frac{1}{2} \sum_{k > j} \frac{h_{\tau_k}}{h_3}}.
\end{equation}
Multiplying these out, we find for the total Miura operator
\begin{equation}
\label{miuraconftransf}
\mathcal{L}(z) \to \tilde{\mathcal{L}}(\tilde{z}) = \left( \frac{d\tilde{z}}{dz}\right)^{-\frac{1}{2} - \frac{1}{2} \sum_{k} \frac{h_{\tau_k}}{h_3}} \mathcal{L}(z) \left( \frac{d\tilde{z}}{dz}\right)^{\frac{1}{2} - \frac{1}{2} \sum_{k} \frac{h_{\tau_k}}{h_3}}.
\end{equation}
which generalizes (\ref{totalmiura3}) to the case of $\mathcal{Y}_{N_1,N_2,N_3}$.

\section{R-matrix}
\label{secrmatrix}
We saw that the free field representation of $\mathcal{Y}_{N_1,N_2,N_3}$ obtained from the Miura transformation depends on the way in which we order the free bosons. Any permutation of free bosons can be obtained by composing the elementary transpositions of neighboring factors so we can focus on these. We define $R$-matrix $\mathcal{R}_{j,j+1}$ to be the intertwiner between the free field representations which differ by order of two neighboring free fields:
\begin{equation}
\label{rmatrixdef}
\mathcal{R}^{(\tau_j \tau_{j+1})}_{j,j+1} \mathcal{L}_j^{(\tau_j)}(z) \mathcal{L}_{j+1}^{(\tau_{j+1})}(z) = \mathcal{L}_{j+1}^{(\tau_{j+1})}(z) \mathcal{L}_j^{(\tau_j)}(z) \mathcal{R}^{(\tau_j \tau_{j+1})}_{j,j+1}
\end{equation}
The operator $\mathcal{R}_{j,j+1}$ defined in this way acts only on Fock spaces $\mathcal{F}_j$ and $\mathcal{F}_{j+1}$, i.e. it is a linear map
\begin{equation}
\mathcal{R}_{j,j+1}: \mathcal{F}_j \otimes \mathcal{F}_{j+1} \to \mathcal{F}_j \otimes \mathcal{F}_{j+1}.
\end{equation}

\paragraph{R-matrix for two bosons of third type}
The simplest situation is if we consider two $R$-matrices of $3$rd type. In this case the equation (\ref{rmatrixdef}) reads
\begin{equation}
\mathcal{R}^{(33)}_{12} (\alpha_0 \partial + J_1(z)) (\alpha_0 \partial + J_2(z)) = (\alpha_0 \partial + J_2(z)) (\alpha_0 \partial + J_1(z)) \mathcal{R}^{(33)}_{12}
\end{equation}
or
\begin{multline}
\mathcal{R}^{(33)}_{12} \left[ \alpha_0^2 \partial^2 + (J_1(z) + J_2(z)) \alpha_0 \partial + (J_1 J_2)(z) + \alpha_0 (\partial J_2)(z) \right] = \\ = \left[ \alpha_0^2 \partial^2 + (J_1(z) + J_2(z)) \alpha_0 \partial + (J_1 J_2)(z) + \alpha_0 (\partial J_1)(z) \right] \mathcal{R}^{(33)}_{12}
\end{multline}
Comparing the terms of different powers of $\partial$, we see that
\begin{align}
\mathcal{R}_{12}^{(33)} \left[J_1(z) + J_2(z)\right] & = \left[J_1(z) + J_2(z)\right] \mathcal{R}_{12}^{(33)} \\
\mathcal{R}_{12}^{(33)} \left[(J_1 J_2)(z) + \alpha_0 \partial J_2(z)\right] & = \left[(J_1 J_2)(z) + \alpha_0 \partial J_1(z)\right] \mathcal{R}_{12}^{(33)}
\end{align}
The first of these equations implies that $R_{12}$ commutes with the total current $J_1+J_2$. For this reason, it is useful to rewrite these equations in terms of orthogonal combinations $J_+ \equiv J_1 + J_2$ and $J_- \equiv h_3 (J_1 - J_2)$. Commutativity with $J_+$ implies that $R_{12}$ is constructed from mode operators of $J_-$ only and furthermore it needs to satisfy
\begin{equation}
\label{r33req}
\mathcal{R}_{12}^{(33)} \left[(J_- J_-)(z) - \frac{2h_3^3}{h_1 h_2 h_3} \partial J_-(z) \right] = \left[(J_- J_-)(z) + \frac{2h_3^3}{h_1 h_2 h_3} \partial J_-(z) \right] \mathcal{R}_{12}^{(33)}.
\end{equation}
Here $J_-(z)$ is a current which satisfies OPE
\begin{equation}
J_-(z) J_-(w) \sim -\frac{2 h_3^3}{h_1 h_2 h_3} \frac{1}{(z-w)^2} \sim -\frac{2\lambda_1\lambda_2}{\lambda_3^2} \frac{1}{(z-w)^2}.
\end{equation}
Note that the requirement (\ref{r33req}) is closely related to Liouville reflection operator \cite{zamolodchikov2007lectures,Zhu:2015nha}.

\paragraph{R-matrix for two bosons of 1st and 2nd type}
The situation with bosons of first or second type is analogous, but now the operator $\mathcal{L}^{(\tau)}(z)$ ($\tau=1,2$) is pseudo-differential operator with infinite number of derivatives so it would seem that we get more constraints on $R^{(\tau\tau)}$. But it turns out (as we expect from the triality symmetry of the three bosonic representations as well as from the fact that with two bosons we are still studying Virasoro algebra) that there are again only two independent conditions. Introducing again currents $J_{\pm}$, the definition of $R^{(\tau\tau)}$ reduces to equations
\begin{align}
\mathcal{R}^{(\tau\tau)}_{12} J_+(z) & = J_+(z) \mathcal{R}^{(\tau\tau)}_{12} \\
\label{rfielddef}
\mathcal{R}^{(\tau\tau)}_{12} \left[ (J_- J_-)(z) - \frac{2h_\tau^3}{h_1 h_2 h_3} (\partial J_-)(z) \right] & = \left[ (J_- J_-)(z) + \frac{2h_\tau^3}{h_1 h_2 h_3} (\partial J_-)(z) \right] \mathcal{R}^{(\tau\tau)}_{12}
\end{align}
where
\begin{equation}
J_-(z) J_-(w) \sim -\frac{2h_\tau^3}{h_1 h_2 h_3} \frac{1}{(z-w)^2}.
\end{equation}
These take exactly the same form as the equations for $R^{(33)}$. Note that the Miura transformation was defined in a way which is not symmetric with respect to the three directions, but the definitions of Fock space $R$-matrices that we found are completely symmetric.

\subsection{R-matrices of the mixed type}
We can also consider what happens if we exchange the two bosons of different types. Consider for example
\begin{equation}
\label{r12def}
\mathcal{R}_{12}^{(12)} \mathcal{L}_1^{(1)}(z) \mathcal{L}_2^{(2)}(z) = \mathcal{L}_2^{(2)}(z) \mathcal{L}_1^{(1)}(z) \mathcal{R}_{12}^{(12)}.
\end{equation}
We get a set of equations by comparing the coefficients of various derivatives. The leading equation is
\begin{equation}
\mathcal{R}_{12}^{(12)} (J_1+J_2)(z) = (J_1+J_2)(z) \mathcal{R}_{12}^{(12)}
\end{equation}
so $\mathcal{R}_{12}^{(12)}$ commutes with all mode operators of $J_+ = J_1+J_2$. The linear combination of $J_1$ and $J_2$ orthogonal to $J_+$ is this time
\begin{equation}
J_-(z) = h_2 J_1(z) - h_1 J_2(z)
\end{equation}
and it has OPE with itself
\begin{equation}
J_-(z) J_-(w) \sim \frac{1}{(z-w)^2}.
\end{equation}
The coefficient of the subleading power of $\partial$ in (\ref{r12def}) is
\begin{equation}
\label{rfielddefm}
\mathcal{R}_{12}^{(12)} \left[ (J_- J_-)(z) + \partial J_-(z) \right] = \left[ (J_- J_-)(z) - \partial J_-(z) \right] \mathcal{R}_{12}^{(12)}.
\end{equation}
Unlike in the case of $\mathcal{R}^{\tau\tau}$ where the associated algebra was $\mathcal{Y}_{002}$, i.e. $\hat{\mathfrak{u}}(1) \times \mathfrak{Vir}$, now the algebra obtained from Miura transformation is of the type $\mathcal{Y}_{110}$, i.e. $\hat{\mathfrak{u}}(1)$ times the parafermion algebra. The parafermion algebra is not generated by the stress-energy tensor alone, so by looking at coefficients of lower derivatives in (\ref{r12def}) we find other conditions like
\begin{equation}
\mathcal{R}_{12}^{(12)} \left[ 4(J_- (J_- J_-)) + 6(\partial J_- J_-) + \partial^2 J_- \right] = \left[ 4(J_- (J_- J_-)) - 6(\partial J_- J_-) + \partial^2 J_- \right] \mathcal{R}_{12}^{(12)}
\end{equation}
at dimension $3$ or
\begin{multline}
\mathcal{R}_{12}^{(12)} \left[ 3(J_- (J_- (J_- J_-))) + 6 (\partial J_- (J_- J_-)) + 3(\partial J_- \partial J_-) + 6 (\partial^2 J_- J_-) + 2\partial^3 J_- \right] = \\
= \left[ 3(J_- (J_- (J_- J_-))) - 6 (\partial J_- (J_- J_-)) + 3(\partial J_- \partial J_-) + 6 (\partial^2 J_- J_-) - 2\partial^3 J_- \right] \mathcal{R}_{12}^{(12)}
\end{multline}
at dimension $4$. We can find higher order relations either by studying coefficients of lower derivatives in (\ref{r12def}) or by studying the operator product expansions of relations that we already found. Since the primaries of dimension $4$ and higher in $\mathcal{Y}_{110}$ appear in the OPE of spin $3$ field itself, the higher order equations for $\mathcal{R}$-matrix will be satisfied if they are satisfied for fields of spin $1$, $2$ and $3$.

\subsection{Mode expansions}
Let us now study the mode expansions of (\ref{rfielddef}) and (\ref{rfielddefm}). First of all, notice that all of these equations can be compactly written as
\begin{equation}
\label{rdefrho}
\mathcal{R} \left[ (J_- J_-)(z) + \rho \partial J_-(z) \right] = \left[ (J_- J_-)(z) - \rho \partial J_-(z) \right] \mathcal{R}
\end{equation}
with normalization
\begin{equation}
J_-(z) J_-(w) \sim \frac{\rho}{(z-w)^2}
\end{equation}
where $\rho$ takes the value
\begin{equation}
\rho = -\frac{h_\sigma h_\tau(h_\sigma+h_\tau)}{\sigma_3}
\end{equation}
for $\mathcal{R}^{\tau\sigma}$. Note that the field
\begin{equation}
\frac{1}{2\rho} \left( (J_- J_-)(z) \pm \rho \partial J_-(z) \right)
\end{equation}
satisfies OPE of stress-energy tensor with the central charge $c = 1 - 3\rho$.

\paragraph{Complex plane}
Let us consider the mode expansion on the complex plane. We have
\begin{equation}
J_j(z) = \sum_{j \in \mathbbm{Z}} \frac{a_{j,m}}{z^{m+1}}.
\end{equation}
The operator product expansion of $J_j(z)$ with itself can be translated to commutation relations
\begin{equation}
\left[ a_{j,m}, a_{k,n} \right] = -\frac{h_{\tau_j}}{h_1 h_2 h_3} m \delta_{m+n,0} \delta_{jk}
\end{equation}
and analogously for modes of $J_{\pm}$. Note that the zero modes $a_{j,0}$ are central, i.e. commute with any other mode operator. The mode expansion of (\ref{rdefrho}) is
\begin{equation}
\label{rexpplane}
\mathcal{R} \left[ \sum_{k \in \mathbbm{Z}} : a^-_k a^-_{m-k} : - \rho (m+1) a^-_m \right] = \left[ \sum_{k \in \mathbbm{Z}} : a^-_k a^-_{m-k} : + \rho (m+1) a^-_m \right] \mathcal{R}
\end{equation}

\paragraph{Cylinder}
In the existing literature \cite{Zhu:2015nha}, the mode expansions are usually done on the cylinder which is related to the complex plane by the exponential mapping. Denoting the cylinder coordinate by $\tilde{z}$, the map is
\begin{equation}
\label{cylindertoplane}
z = e^{\tilde{z}}.
\end{equation}
We must remember that as discussed in section \ref{secconftrans}, under conformal transformations $J_-(z)$ does not transform as a primary field, but instead has an anomalous transformation
\begin{equation}
\tilde{J}_-(\tilde{z}) = \left(\frac{d \tilde{z}}{dz}\right)^{-1} J_-(z) + \frac{\rho}{2} \left( \frac{d\tilde{z}}{dz}\right)^{-2} \left( \frac{d^2 \tilde{z}}{dz^2} \right).
\end{equation}
In terms of modes, this means
\begin{equation}
\tilde{J}_-(\tilde{z}) = \sum_{k \in \mathbbm{Z}} a^-_k e^{-k\tilde{z}} - \frac{\rho}{2}.
\end{equation}
This is a usual mode expansion of free boson on the cylinder, apart from the fact that the zero mode is shifted by a constant,
\begin{equation}
\label{cyltoplshift}
a^-_{0(cyl)} = a^-_{0(pl)} - \frac{\rho}{2}.
\end{equation}
Using this in equation (\ref{rexpplane}), we can write an analogous formula on the cylinder,
\begin{equation}
\label{rexpcyl}
\mathcal{R} \left[ \sum_{k \in \mathbbm{Z}} : a^-_{k(cyl)} a^-_{m-k(cyl)} : - \rho m a^-_{m(cyl)} \right] = \left[ \sum_{k \in \mathbbm{Z}} : a^-_{k(cyl)} a^-_{m-k(cyl)} : + \rho m a^-_{m(cyl)} \right] \mathcal{R}
\end{equation}
which is the form used in \cite{Zhu:2015nha}. Note that both of the sides of this equation come from free bosons considered in different ordering, so effectively we should use the opposite transformation rule between the modes on the RHS. But conjugation by $\mathcal{R}$ that appears in this equation exactly compensates for this, i.e. conjugation of operator implementing the conformal transformation in one ordering by $\mathcal{R}$ changes it into conformal transformation in the other ordering.

The equation (\ref{rexpcyl}) could have been obtained by directly taking the mode expansion of equation (\ref{rdefrho}) on the cylinder. The independence of the Fock $\mathcal{R}$-matrix on the conformal frame is a consequence of the transformation property (\ref{miuraconftransf}) of Miura operator under conformal transformations.

In the case of $\mathcal{R}$-matrix of mixed type, we should consider not only spin $2$ defining relation (\ref{rexpcyl}) but also spin $3$ relation. On cylinder, this requires
\begin{equation}
\left[ \mathcal{R}, 4 a_0^3 - a_0 + 24 a_0 \sum_{k>0} : a_{-k} a_k + 12 \sum_{j,k>0} (a_{-j} a_{-k} a_{j+k} + a_{-j-k} a_j a_k) \right] = 0
\end{equation}
for mode number zero and
\begin{multline}
\label{spin3primaryexp}
\mathcal{R} \left[ 4 \sum_{j+k+l=m} :a_j a_k a_l: - 6 \sum_j j :a_j a_{m-j}: + (m^2-1) a_m \right] \\
= \left[ 4 \sum_{j+k+l=m} :a_j a_k a_l: + 6 \sum_j j :a_j a_{m-j}: + (m^2-1) a_m \right] \mathcal{R}
\end{multline}
for $m \neq 0$. An easy way to find these mode expansions is to start in the plane and use (\ref{cyltoplshift}) to transform to cylinder. As consistency check, it is also easy to verify that the mode expansions (\ref{spin3primaryexp}) are those of spin $3$ primary field so it is enough to verify that the zero mode transforms correctly under the conjugation by $\mathcal{R}$-matrix.

\subsection{Expansion of $\mathcal{R}$ at large spectral parameter}
\label{secRuexp}
In \cite{Zhu:2015nha} an expansion of (\ref{rexpcyl}) at infinite value of the central element $a^-_{0(cyl)}$ was studied. In the rest of this section, we will write simply $a_m$ instead of $a^-_{m(cyl)}$. The $\mathcal{R}$-matrix has to satisfy
\begin{equation}
\label{rdefmode}
\mathcal{R} \left[ \sum^\prime_{k \in \mathbbm{Z}} a_k a_{m-k} + 2a_0 a_m - \rho m a_m \right] = \left[ \sum^\prime_{k \in \mathbbm{Z}} a_k a_{m-k} + 2a_0 a_m + \rho m a_m \right] \mathcal{R}
\end{equation}
for $m \neq 0$ where $\sum^\prime$ means that we leave out the terms which contain the zero mode $a_0$. For $m=0$ we have
\begin{equation}
\mathcal{R} \left[ \sum_{k>0} a_{-k} a_k + \frac{a_0^2}{2} \right] = \left[ \sum_{k>0} a_{-k} a_k + \frac{a_0^2}{2} \right] \mathcal{R},
\end{equation}
i.e. $\mathcal{R}$ must preserve the Fock level. For large values of $a_0$ the $\rho$-dependence drops out so we can look for expansion
\begin{equation}
\mathcal{R} = R^{(0)} + a_0^{-1} R^{(1)} + a_0^{-2} R^{(2)} + \ldots
\end{equation}
with $\mathcal{R}^{(0)} = \mathbbm{1}$. The $j$th order equation we have to solve is
\begin{equation}
2 \left[ R^{(j)}, a_m \right] = \rho m \left\{ R^{(j-1)}, a_m \right\} - \left[ R^{(j-1)}, \sum_{k \in \mathbbm{Z}}^\prime a_k a_{m-k} \right]
\end{equation}
which for $j=1$ simplifies to
\begin{equation}
\left[ R^{(1)}, a_m \right] = \rho m a_m.
\end{equation}
The constant term is not determined by this equation, so that we can fix it so that $\mathcal{R}^{(1)}$ annihilates the highest weight state. Under this condition we find
\begin{equation}
R^{(1)} = -\sum_{k>0} a_{-k} a_k.
\end{equation}
At next order, we have
\begin{equation}
2 \left[ R^{(2)}, a_m \right] = \rho m \left\{ \mathcal{R}^{(1)}, a_m \right\} - \left[ \mathcal{R}^{(1)}, \sum_{k \in \mathbbm{Z}}^\prime a_k a_{m-k} \right]
\end{equation}
The solution of this equation annihilating the highest weight state is
\begin{equation}
R^{(2)} = \frac{1}{2} \sum_{j_1,j_2>0} \left(a_{-j_1-j_2} a_{j_1} a_{j_2} + a_{-j_1} a_{-j_2} a_{j_1+j_2} \right) + \frac{1}{2} R^{(1)} R^{(1)}.
\end{equation}
Perhaps surprisingly, $R^{(2)}$ is equal to a zero mode of a local field plus a quadratic term in $R^{(1)}$ which is not a zero mode of a density, but can be eliminated by taking a logarithm of $\mathcal{R}$. For this reason, it is convenient following \cite{Zhu:2015nha} to parametrize $\mathcal{R}$ as
\begin{equation}
\mathcal{R} = \exp \left[ a_0^{-1} r^{(1)} + a_0^{-2} r^{(2)} + a_0^{-3} r^{(3)} + \ldots \right].
\end{equation}
Even at higher orders, $r^{(j)}$ will be zero modes of local densities of the current $J^{(cyl)}_-(z)-a_0$. At third order, we find
\begin{equation}
R^{(3)} = r^{(3)} + \frac{1}{6} r^{(1)} r^{(1)} r^{(1)} + \frac{1}{2} r^{(1)} r^{(2)} + \frac{1}{2} r^{(2)} r^{(1)}
\end{equation}
with
\begin{align}
r^{(3)} & = -\frac{1}{3} \sum_{j_1,j_2,j_3>0} \left(a_{-j_1-j_2-j_3} a_{j_1} a_{j_2} a_{j_3} + a_{-j_1} a_{-j_2} a_{-j_3} a_{j_1+j_2+j_3} \right) \\
& - \frac{1}{2} \sum_{j_1+j_2=k_1+k_2} a_{-j_1} a_{-j_2} a_{k_1} a_{k_2} + \frac{\rho}{12} \sum_{j>0} a_{-j} a_j - \frac{\rho(\rho+1)}{12} \sum_{j>0} j^2 a_{-j} a_j.
\end{align}
The fourth and fifth order expressions are given in the appendix \ref{apprmat}. Unlike the expansion coefficients of the $R$-matrix itself, the coefficients in expansion of its logarithm can always be expressed at zero modes of local fields. We have
\begin{align}
r^{(1)} = & \, -\frac{1}{2} (J_- J_-)_0 \\
r^{(2)} = & \, \frac{1}{6} (J_- (J_- J_-))_0 \\
r^{(3)} = & \, \frac{1}{12} (J_- (J_- (J_- J_-)))_0 + \frac{\rho}{24} (J_- J_-)_0 + \frac{\rho(\rho+1)}{24} (\partial J_- \partial J_-)_0
\end{align}
where we always leave out the zero mode. \footnote{A. Litvinov considered the expansion of the logarithm of $\mathcal{R}$-matrix to order $u^{-8}$ and all the terms appearing there are zero modes of local currents.}

\paragraph{Expressions for $R$}
As was already discussed, we looked for the large spectral parameter expansion of the logarithm of $R$-matrix which leads to expressions involving zero modes of local fields. But in the following it will also be useful to have expressions for the expansion coefficients of $R$-matrix itself. We will write them in the normal-ordered form which will be convenient later. The first coefficient $r^{(1)}$ is unchanged. For the second one we have
\begin{align}
\nonumber
R^{(2)} & = r^{(2)} + \frac{1}{2}r^{(1)2} = \frac{1}{2} \sum_{j,k>0} (a_{-j-k} a_j a_k + a_{-j} a_{-k} a_{j+k}) + \frac{1}{2} \sum_{j,k>0} a_{-j} a_j a_{-k} a_k \\
& = \frac{1}{2} \sum_{j,k>0} (a_{-j-k} a_j a_k + a_{-j} a_{-k} a_{j+k} + a_{-j} a_{-k} a_j a_k) + \frac{\rho}{2} \sum_{j>0} j a_{-j} a_j \\
\nonumber
& = \frac{1}{2} \sum_{j,k>0} (a_{-j-k} + a_{-j} a_{-k}) (a_{j+k} + a_j a_k) + \frac{1}{2} \sum_{j>0} a_{-j} a_j + \frac{\rho-1}{2} \sum_{j>0} j a_{-j} a_j.
\end{align}
and for the third
\begin{align}
\nonumber
R^{(3)} = & -\frac{1}{6} \sum_{j,k,l>0} a_{-j} a_{-k} a_{-l} a_j a_k a_l - \frac{1}{3} \sum_{j,k,l>0} (a_{-j} a_{-k} a_{-l} a_{j+k+l} + a_{-j-k-l} a_j a_k a_l) \\
\nonumber
& - \frac{1}{2} \sum_{j,k,l>0} (a_{-j} a_{-k} a_{-l} a_{j+k} a_l + a_{-j-k} a_{-l} a_j a_k a_l) - \frac{1}{2} \sum_{j+k=l+m} a_{-j} a_{-k} a_l a_m \\
& - \frac{\rho}{4} \sum_{j,k>0} (j+k) a_{-j} a_{-k} a_j a_k - \frac{\rho}{2} \sum_{j,k>0} (j+k) (a_{-j} a_{-k} a_{j+k} + a_{-j-k} a_j a_k) \\
\nonumber
& + \frac{\rho}{12} \sum_{j>0} a_{-j} a_j - \frac{\rho(1+3\rho)}{12} \sum_{j>0} j^2 a_{-j} a_j.
\end{align}
We see that the expression for $R^{(j)}$ are more complicated than those for $r^{(j)}$ at the same order of the expansion. There are also very few cancellations when going from $r^{(j)}$ to $R^{(j)}$ so indeed taking the logarithm simplifies the large spectral parameter expansion of $R$ considerably. But for $\rho = 1$ these expansion coefficients for $\mathcal{R}$ factorize up to a combination of lower order terms. We have for example
\begin{align}
\nonumber
R^{(3)} = & -\frac{1}{6} \sum_{j,k,l>0} (a_{-j} a_{-k} a_{-l} + a_{-j-k}a_{-l} + a_{-j-l}a_{-k} + a_{-k-l}a_{-j} + 2a_{-j-k-l}) \times \\
\nonumber
& \times (a_j a_k a_l + a_{j+k}a_l +a_{j+l}a_k + a_{k+l}a_j + 2a_{j+k+l}) \\
& - \sum_{j,k>0} (a_{-j} a_{-k} + a_{-j-k}) (a_j a_k + a_{j+k}) - \frac{1}{4} \sum_{j>0} a_{-j} a_j \\
\nonumber
& - \frac{\rho-1}{4} \sum_{j,k>0} (j+k) a_{-j} a_{-k} a_j a_k - \frac{\rho-1}{2} \sum_{j,k>0} (j+k) (a_{-j} a_{-k} a_{j+k} + a_{-j-k} a_j a_k) \\
\nonumber
& + \frac{\rho-1}{12} \sum_{j>0} \left[1-(3\rho+4)j^2\right] a_{-j} a_j
\end{align}
so at $\rho=1$ the $R$-matrix simplifies considerably,
\begin{align}
\label{rmixexp}
\nonumber
\mathcal{R}(\rho=1) & = \mathbbm{1} + \left( - \frac{1}{a_0} + \frac{1}{2a_0^2} - \frac{1}{4a_0^3} + \ldots \right) \sum_{j_1>0} \mathcal{M}_{-j_1} \mathcal{M}_{j_1} \\
& + \left( \frac{1}{2a_0^2} - \frac{1}{a_0^3} + \ldots \right) \sum_{j_1,j_2>0} \mathcal{M}_{-j_1,-j_2} \mathcal{M}_{j_1,j_2} \\
\nonumber
& - \frac{1}{6a_0^3} \sum_{j_1,j_2,j_3>0} \mathcal{M}_{-j_1,-j_2,-j_3} \mathcal{M}_{j_1,j_2,j_3} + \ldots
\end{align}
with
\begin{align}
\nonumber
\mathcal{M}_{j_1} & = a_{j_1} \\
\mathcal{M}_{j_1,j_2} & = a_{j_1+j_2} + a_{j_1} a_{j_2} \\
\nonumber
\mathcal{M}_{j_1,j_2,j_3} & = a_{j_1}a_{j_2}a_{j_3} + a_{j_1+j_2}a_{j_3} + a_{j_1+j_2}a_{j_3} + a_{j_1+j_2}a_{j_3} + 2a_{j_1+j_2+j_3}
\end{align}
We will see in the following that such a factorization at $\rho=1$ is a general feature of the $\mathcal{R}$-matrix.

\subsection{Matrix elements of $\mathcal{R}$ between simple states}
\label{secrmatrixel}
Although the discussion of the previous section can in principle be continued to higher orders, it is useful in the following to know the matrix elements of $\mathcal{R}$ at least between simple states exactly for all values of the spectral parameter. We still consider the subspace of the Fock space spanned by oscillators of $J_-$. By our choice of normalization of $\mathcal{R}$ we know that
\begin{equation}
\mathcal{R} \ket{0^-} = \ket{0^-}.
\end{equation}
We can now act on this state with (\ref{rdefmode}) with the choice $m=-1$ and we find
\begin{equation}
(2a_0+\rho) \mathcal{R} a^-_{-1} \ket{0^-} = (2a^-_0-\rho) a^-_{-1} \ket{0^-}
\end{equation}
so
\begin{equation}
\mathcal{R} a^-_{-1} \ket{0^-} = \left( \frac{2a^-_0-\rho}{2a^-_0+\rho}\right) a^-_{-1} \ket{0^-}.
\end{equation}
Taking a logarithm of $\mathcal{R}$ we find
\begin{equation}
\log \mathcal{R} a^-_{-1} \ket{0^-} = \left( -\frac{\rho}{a^-_0} -\frac{\rho^3}{12(a^-_0)^3} -\frac{\rho^5}{80(a^-_0)^5} -\frac{\rho^7}{448(a^-_0)^7} + \ldots \right) a^-_{-1} \ket{0^-}
\end{equation}
which reproduces the results of the previous section. Dually, we have
\begin{equation}
\bra{0^-} a^-_1 \mathcal{R} = \left(\frac{2a^-_0-\rho}{2a^-_0+\rho}\right) \bra{0^-} a^-_1.
\end{equation}
At the next level, we can act twice with $m=-1$ (\ref{rdefmode}) or once with $m=-2$ on the vacuum. We find
\begin{align}
\label{ractlevel2}
\mathcal{R} a^-_{-2} \ket{0^-} = & \frac{4(a^-_0)^3 - 2a^-_0 \rho + \rho^2 - 3a^-_0 \rho^2 - \rho^3}{(2a^-_0+\rho)(2(a^-_0)^2-\rho+3(a^-_0) \rho+\rho^2)} a^-_{-2} \ket{0^-} \\
& + \frac{4a^-_0 \rho}{(2a^-_0+\rho)(2(a^-_0)^2-\rho+3a^-_0 \rho+\rho^2)} (a^-_{-1})^2 \ket{0^-} \\
\mathcal{R} (a^-_{-1})^2 \ket{0^-} = & \frac{4a^-_0 \rho^2}{(2a^-_0+\rho)(2(a^-_0)^2-\rho+3a^-_0 \rho+\rho^2)} a^-_{-2} \ket{0^-} \\
& + \frac{4(a^-_0)^3-2a^-_0 \rho-\rho^2-3a^-_0 \rho^2+\rho^3}{(2a^-_0+\rho)(2(a^-_0)^2-\rho+3a^-_0 \rho+\rho^2)} (a^-_{-1})^2 \ket{0^-}.
\end{align}
Proceeding in similar way, we can determine in principle all the matrix elements of $\mathcal{R}$ exactly without using any Taylor expansion or without knowledge of Fock-space formula for $\mathcal{R}$. Requiring $\mathcal{R} \ket{0} = \ket{0}$ uniquely fixes $\mathcal{R}$.

\subsection{Yang-Baxter equation}
If we consider three bosonic Fock spaces, we can derive an important equation satisfied by the $R$-matrices which is the Yang-Baxter equation. Namely, starting from $\mathcal{L}_3 \mathcal{L}_2 \mathcal{L}_1$
we can bring it into the opposite order in two different ways:
\begin{align}
\nonumber
\mathcal{L}_3 \mathcal{L}_2 \mathcal{L}_1 = {} & \mathcal{R}_{23} \mathcal{L}_2 \mathcal{L}_3 \mathcal{L}_1 \mathcal{R}_{23}^{-1} = \mathcal{R}_{23} \mathcal{R}_{13} \mathcal{L}_2 \mathcal{L}_1 \mathcal{L}_3 \mathcal{R}_{13}^{-1} \mathcal{R}_{23}^{-1} = \mathcal{R}_{23} \mathcal{R}_{13} \mathcal{R}_{12} \mathcal{L}_1 \mathcal{L}_2 \mathcal{L}_3 \mathcal{R}_{12}^{-1} \mathcal{R}_{13}^{-1} \mathcal{R}_{23}^{-1} \\
= {} & \mathcal{R}_{12} \mathcal{L}_3 \mathcal{L}_1 \mathcal{L}_2 \mathcal{R}_{12}^{-1} = \mathcal{R}_{12} \mathcal{R}_{13} \mathcal{L}_1 \mathcal{L}_3 \mathcal{L}_2 \mathcal{R}_{13}^{-1} \mathcal{R}_{12}^{-1} = \mathcal{R}_{12} \mathcal{R}_{13} \mathcal{R}_{23} \mathcal{L}_1 \mathcal{L}_2 \mathcal{L}_3 \mathcal{R}_{23}^{-1} \mathcal{R}_{13}^{-1} \mathcal{R}_{12}^{-1}
\end{align}
The Fock space operator completely exchanging the order is thus
\begin{equation}
\mathcal{R}_{12} \mathcal{R}_{13} \mathcal{R}_{23} = \mathcal{R}_{23} \mathcal{R}_{13} \mathcal{R}_{12}
\end{equation}
which is the celebrated Yang-Baxter equation. Note that each $\mathcal{R}$ depends on the types $\tau_j$ of the Fock spaces on which it acts. Also, as always the zero mode of the $\hat{\mathfrak{u}}(1)$ currents plays the role of the spectral parameter and since $\mathcal{R}_{jk}$ depends (up to a conventional rescaling) only on the difference of zero modes, what we find is exactly the Yang-Baxter equation with an additive spectral parameter (i.e. $R$-matrix of the rational type).

As usual in the algebraic Bethe ansatz \cite{Faddeev:1996iy,Nepomechie:1998jf}, once we have a solution of Yang-Baxter equation for certain choice of representation spaces, we can take their tensor products and the corresponding products of $R$-matrices again satisfy the Yang-Baxter equation with more complicated representation spaces. Consider $N$ Fock spaces $\mathcal{F}_j$ and one additional auxiliary space $\mathcal{F}_A$. We will call $\mathcal{F}_Q \equiv \mathcal{F}_1 \otimes \ldots \otimes \mathcal{F}_N$ the quantum space. We can use $\mathcal{F}_A$ to define a monodromy operator
\begin{equation}
\mathcal{T}_A \equiv \mathcal{R}_{A1} \mathcal{R}_{A2} \cdots \mathcal{R}_{AN}.
\end{equation}
Considering another auxiliary space $\mathcal{F}_B$ (but the same quantum space) and using the Yang-Baxter equation, it is easy to see that
\begin{equation}
\label{monoybe}
\mathcal{R}_{AB} \mathcal{T}_A \mathcal{T}_B = \mathcal{T}_B \mathcal{T}_A \mathcal{R}_{AB},
\end{equation}
i.e. we see that $\mathcal{T}_A$ satisfies again the Yang-Baxter equation. Since $\mathcal{R}_{AB}$ is generically invertible, exchanging the order of $\mathcal{T}_A$ and $\mathcal{T}_B$ (which are still operators in the quantum space) can be undone by a similarity transformation $\mathcal{R}_{AB}$ in the auxiliary space. If the representation spaces $\mathcal{F}_A$ and $\mathcal{F}_B$ were finite-dimensional, we could simply take the trace of $\mathcal{T}_A$ over $\mathcal{F}_A$ to find an infinite set of commuting operators in the quantum space $\mathcal{F}_1 \otimes \mathcal{F}_2 \otimes \cdots \otimes \mathcal{F}_N$ \cite{Faddeev:1996iy,Nepomechie:1998jf}. In our application we cannot do this because the trace over infinite-dimensional vector space does not need to converge \footnote{We could regularize the trace by including a Boltzmann factor to ensure convergence \cite{Feigin:2015raa}. From the CFT point of view this would lead to correlation functions evaluated on a torus rather than on a cylinder.}, but we can use a special property of $\mathcal{R}_{AB}$ which is the fact that it preserves the Fock vacuum vector,
\begin{equation}
\mathcal{R}_{AB} \ket{0}_A \otimes \ket{0}_B = \ket{0}_A \otimes \ket{0}_B
\end{equation}
and
\begin{equation}
\bra{0}_A \otimes \bra{0}_B \mathcal{R}_{AB} = \bra{0}_A \otimes \bra{0}_B.
\end{equation}
We can thus define the (analogue of) transfer matrix (here $\tau$ is the type of the auxiliary Fock space)
\begin{equation}
\label{hoperator}
\mathcal{H}^{(\tau)} \equiv \bra{0}_A \mathcal{T}^{(\tau)}_A \ket{0}_A
\end{equation}
which is an operator on the quantum space $\mathcal{F}_1 \otimes \cdots \otimes \mathcal{F}_N$ which moreover depends on the spectral parameter $a_{A,0}$. Taking the vacuum-to-vacuum matrix element of (\ref{monoybe}) we find
\begin{align}
\nonumber
\mathcal{H}(a_{A,0}) \mathcal{H}(a_{B,0}) & = \bra{0}_A \bra{0}_B \mathcal{T}_A \mathcal{T}_B \ket{0}_A \ket{0}_B = \bra{0}_A \bra{0}_B \mathcal{R}_{AB} \mathcal{T}_A \mathcal{T}_B \ket{0}_A \ket{0}_B \\
& = \bra{0}_A \bra{0}_B \mathcal{T}_B \mathcal{T}_A \mathcal{R}_{AB} \ket{0}_A \ket{0}_B = \bra{0}_A \bra{0}_B \mathcal{T}_B \mathcal{T}_A \ket{0}_A \ket{0}_B \\
\nonumber
& = \mathcal{H}(a_{B,0}) \mathcal{H}(a_{A,0}).
\end{align}
We have thus constructed an infinite set of operators acting in the quantum space which commute among themselves (since the previous equation is true for arbitrary values of the spectral parameter).

\subsection{Yangian generators}
It is well-known from the algebraic Bethe ansatz approach to XXX spin chains that the transfer matrix is not the only interesting object that can be constructed from the monodromy matrix. In fact, for $GL(N)$ spin chain one can consider all $N^2$ matrix elements of $\mathcal{R}$ in the $N$-dimensional auxiliary space and the associated quantum space operators satisfy the defining relations of Yangian of $\mathfrak{gl}(N)$.

In our situation the auxiliary space is the Fock space of one free boson which is infinite dimensional, the states being conveniently labeled by the Young diagrams. The associated Yangian of $\widehat{\mathfrak{gl}}(1)$ would be thus given by an infinite set of generators and relations. For our purposes, it will be enough to first consider the first three of them. We already defined one of them, the analogue of the transfer matrix (\ref{hoperator}). It was defined as the vacuum-to-vacuum matrix element of $\mathcal{T}$ in the auxiliary space. Analogously we can introduce the matrix elements
\begin{equation}
\mathcal{E}^{(\tau)} = \bra{0}_A \mathcal{T}^{(\tau)}_A a_{A,-1} \ket{0}_A
\end{equation}
and
\begin{equation}
\mathcal{F}^{(\tau)} = \bra{0}_A a_{A,1} \mathcal{T}^{(\tau)}_A \ket{0}_A
\end{equation}
As we will shortly see, these will be the simplest creation and annihilation operators of our Yangian. Without any assumptions on the details of the quantum space, we can find the first few commutation relations satisfied by these operators. For that we only need the $R$-matrix in auxiliary space. We have already shown that $\mathcal{H}^{(\tau)}$ commute among themselves (for any value of the spectral parameter) but in fact the same argument shows that
\begin{equation}
\left[ \mathcal{H}^{(\tau_A)}(a_{A,0}), \mathcal{H}^{(\tau_B)}(a_{B,0}) \right] = 0,
\end{equation}
i.e. the operators $\mathcal{H}$ commute even for a different choice of the type of the auxiliary Fock space. To find the commutation relation between $\mathcal{E}$ and $\mathcal{H}$ we write
\begin{equation}
\mathcal{H}^{(\tau_A)} \mathcal{E}^{(\tau_B)} = \bra{0}_A \bra{0}_B \mathcal{T}_A \mathcal{T}_B \, a_{B,-1} \ket{0}_A \ket{0}_B = \bra{0}_A \bra{0}_B \mathcal{R}_{AB}^{-1} \mathcal{T}_B \mathcal{T}_A \mathcal{R}_{AB} \, a_{B,-1} \ket{0}_A \ket{0}_B.
\end{equation}
The inverse of $\mathcal{R}_{AB}$ preserves the vacuum on the left just as $\mathcal{R}_{AB}$ did. On the other hand, the action of $\mathcal{R}_{AB}$ on the right on the excited state can be calculated from the results of section \ref{secrmatrixel}:
\begin{align}
\nonumber
\mathcal{R}_{AB} \, a_{A,-1} \ket{0}_{AB} = & \, \frac{u_A-u_B+h_{\tau_A}-h_{\tau_B}}{u_A-u_B+h_{\tau_A}} a_{A,-1} \ket{0}_{AB} +\frac{h_{\tau_A}}{u_A-u_B+h_{\tau_A}} a_{B,-1} \ket{0}_{AB} \\
\mathcal{R}_{AB} \, a_{B,-1} \ket{0}_{AB} = & \, \frac{h_{\tau_B}}{u_A-u_B+h_{\tau_A}} a_{A,-1} \ket{0}_{AB} + \frac{u_A-u_B}{u_A-u_B+h_{\tau_A}} a_{B,-1} \ket{0}_{AB}
\end{align}
where we introduced the spectral parameter
\begin{equation}
\label{spectparama0}
u_A \equiv -\frac{\sigma_3}{h_{\tau_A}} a_{A,0} - \frac{h_{\tau_A}}{2}.
\end{equation}
Note that we have
\begin{equation}
a^-_0 = h_{\tau_B} a_{A,0} - h_{\tau_A} a_{B,0} = -\frac{h_{\tau_A} h_{\tau_B}}{\sigma_3} (u_A - u_B) - \frac{h_{\tau_A} h_{\tau_B}(h_{\tau_A}-h_{\tau_B})}{2\sigma_3}.
\end{equation}
Using this, we find the first non-trivial Yangian commutation relation
\begin{align}
\mathcal{H}^{(\tau_A)}(u_A) \mathcal{E}^{(\tau_B)}(u_B) = & \frac{h_{\tau_B}}{u_A-u_B+h_{\tau_A}} \mathcal{H}^{(\tau_B)}(u_B) \mathcal{E}^{(\tau_A)}(u_A) \\
& + \frac{u_A-u_B}{u_A-u_B+h_{\tau_A}} \mathcal{E}^{(\tau_B)}(u_B) \mathcal{H}^{(\tau_A)}(u_A).
\end{align}
and symmetrically
\begin{align}
\mathcal{E}^{(\tau_A)}(u_A) \mathcal{H}^{(\tau_B)}(u_B) = & \frac{u_A-u_B+h_{\tau_A}-h_{\tau_B}}{u_A-u_B+h_{\tau_A}} \mathcal{H}^{(\tau_B)}(u_B) \mathcal{E}^{(\tau_A)}(u_A) \\
& +\frac{h_{\tau_A}}{u_A-u_B+h_{\tau_A}} \mathcal{E}^{(\tau_B)}(u_B) \mathcal{H}^{(\tau_A)}(u_A).
\end{align}
We can proceed analogously to derive the commutation relation between $\mathcal{H}$ and $\mathcal{F}$ starting from
\begin{align}
\bra{0}_{AB} a_{A,1} \mathcal{R}_{AB}^{-1} & = \frac{u_A-u_B}{u_A-u_B-h_{\tau_B}} \bra{0}_{AB} a_{A,1} + \frac{-h_{\tau_A}}{u_A-u_B-h_{\tau_B}} \bra{0}_{AB} a_{B,1} \\
\bra{0}_{AB} a_{B,1} \mathcal{R}_{AB}^{-1} & = \frac{-h_{\tau_B}}{u_A-u_B-h_{\tau_B}} \bra{0}_{AB} a_{A,1} + \frac{u_A-u_B+h_{\tau_A}-h_{\tau_B}}{u_A-u_B-h_{\tau_B}} \bra{0}_{AB} a_{B,1}
\end{align}
and we find
\begin{align}
\mathcal{H}^{(\tau_A)}(u_A) \mathcal{F}^{(\tau_B)}(u_B) = & -\frac{h_{\tau_B}}{u_A-u_B-h_{\tau_B}} \mathcal{H}^{(\tau_B)}(u_B) \mathcal{F}^{(\tau_A)}(u_A) \\
& + \frac{u_A-u_B+h_{\tau_A}-h_{\tau_B}}{u_A-u_B-h_{\tau_B}} \mathcal{F}^{(\tau_B)}(u_B) \mathcal{H}^{(\tau_A)}(u_A)
\end{align}
and
\begin{align}
\mathcal{F}^{(\tau_A)}(u_A) \mathcal{H}^{(\tau_B)}(u_B) = & \, \frac{u_A-u_B}{u_A-u_B-h_{\tau_B}} \mathcal{H}^{(\tau_B)}(u_B) \mathcal{F}^{(\tau_A)}(u_A) \\
& -\frac{h_{\tau_A}}{u_A-u_B-h_{\tau_B}} \mathcal{E}^{(\tau_B)}(u_B) \mathcal{H}^{(\tau_A)}(u_A).
\end{align}
In particular, if both auxiliary spaces are of the same type, $\tau_A = \tau = \tau_B$, we have more simply
\begin{align}
\label{ybehe2start}
(u-v+h_{\tau}) \mathcal{H}^{(\tau)}(u) \mathcal{E}^{(\tau)}(v) & = h_{\tau} \mathcal{H}^{(\tau)}(v) \mathcal{E}^{(\tau)}(u) + (u-v) \mathcal{E}^{(\tau)}(v) \mathcal{H}^{(\tau)}(u) \\
(u-v+h_{\tau}) \mathcal{E}^{(\tau)}(u) \mathcal{H}^{(\tau)}(v) & = (u-v) \mathcal{H}^{(\tau)}(v) \mathcal{E}^{(\tau)}(u) +h_{\tau} \mathcal{E}^{(\tau)}(v) \mathcal{H}^{(\tau)}(u).
\end{align}
and
\begin{align}
(u-v-h_{\tau}) \mathcal{H}^{(\tau)}(u) \mathcal{F}^{(\tau)}(v) & = -h_{\tau} \mathcal{H}^{(\tau)}(v) \mathcal{F}^{(\tau)}(u) + (u-v) \mathcal{F}^{(\tau)}(v) \mathcal{H}^{(\tau)}(u) \\
\label{ybehe2end}
(u-v-h_{\tau}) \mathcal{F}^{(\tau)}(u) \mathcal{H}^{(\tau)}(v) & = (u-v) \mathcal{H}^{(\tau)}(v) \mathcal{F}^{(\tau)}(u) - h_{\tau} \mathcal{F}^{(\tau)}(v) \mathcal{H}^{(\tau)}(u).
\end{align}
We can think of these relations as the first few relations of the affine Yangian. They are true for any representation of the algebra, i.e. any choice of the quantum space as long as the quantum space is obtained in a way that respects the Yang-Baxter equation. The equations we found can be also rewritten as
\begin{align}
\label{ybehe}
(u-v) \left[ \mathcal{H}^{(\tau)}(u), \mathcal{E}^{(\tau)}(v) \right] & = -h_{\tau} \left( \mathcal{H}^{(\tau)}(u) \mathcal{E}^{(\tau)}(v) - \mathcal{H}^{(\tau)}(v) \mathcal{E}^{(\tau)}(u) \right) \\
(u-v) \left[ \mathcal{E}^{(\tau)}(u), \mathcal{H}^{(\tau)}(v) \right] & = -h_{\tau} \left( \mathcal{E}^{(\tau)}(u) \mathcal{H}^{(\tau)}(v) - \mathcal{E}^{(\tau)}(v) \mathcal{H}^{(\tau)}(u) \right) \\
\label{ybehf}
(u-v) \left[ \mathcal{H}^{(\tau)}(u), \mathcal{F}^{(\tau)}(v) \right] & = -h_{\tau} \left( \mathcal{H}^{(\tau)}(v) \mathcal{F}^{(\tau)}(u) - \mathcal{H}^{(\tau)}(u) \mathcal{F}^{(\tau)}(v) \right) \\
(u-v) \left[ \mathcal{F}^{(\tau)}(u), \mathcal{H}^{(\tau)}(v) \right] & = -h_{\tau} \left( \mathcal{F}^{(\tau)}(v) \mathcal{H}^{(\tau)}(u) - \mathcal{F}^{(\tau)}(u) \mathcal{H}^{(\tau)}(v) \right)
\end{align}
Note that the first two of these equations are not independent: by replacing $u \leftrightarrow v$ in the first equation and eliminating $\mathcal{H}(u)\mathcal{E}(v)$ from the resulting system of linear equations gives the second equation of the list. A similar argument shows the equivalence of the last two equations. Equating the first two and the second two equations we find
\begin{align}
\left[ \mathcal{H}^{(\tau)}(u), \mathcal{E}^{(\tau)}(v) \right] & = \left[ \mathcal{H}^{(\tau)}(v), \mathcal{E}^{(\tau)}(u) \right] \\
\left[ \mathcal{H}^{(\tau)}(v), \mathcal{F}^{(\tau)}(u) \right] & = \left[ \mathcal{H}^{(\tau)}(u), \mathcal{F}^{(\tau)}(v) \right].
\end{align}

\section{Single boson representation}
\label{secboson}
Let us now discuss the results of the previous section more explicitly in the case of a single free boson representation, i.e. the choice $N=1$ (spin chain of length $1$). Let us consider the auxiliary space of type $\tau$ and the quantum space of type $\sigma$. Using the oscillator expansions of $R$ given in section \ref{secRuexp}, we can find the first few terms in the expansion of $\mathcal{H}^{(\tau)}$,
\begin{align}
\nonumber
\mathcal{H}^{(\tau)} = & \, \mathbbm{1} - \frac{h_{\tau}^2}{a^-_0} \sum_{k>0} b_{-k} b_{k} + \frac{1}{2(a^-_0)^2} \Big[ \sum_{j,k>0} (-h_{\tau} b_{-j-k} + h_{\tau}^2 b_{-j} b_{-k}) (-h_{\tau} b_{j+k} + h_{\tau}^2 b_j b_k) \\
& + h_{\tau}^2 \sum_{j>0} b_{-j} b_j + (\rho-1)h_{\tau}^2 \sum_{j>0} j b_{-j} b_j \Big] + \ldots
\end{align}
Here and in the following we denote the oscillators acting in the quantum Fock space by $b_j$. Note that to get from $\mathcal{R}$-matrix to $\mathcal{H}$ we are taking the vacuum-to-vacuum matrix element in the auxiliary space. Furthermore $\mathcal{R}$ only depends on the difference of auxiliary and Fock oscillators and commutes with their sum. If we start from normal-ordered expression for $\mathcal{R}$, this guarantees that $\mathcal{H}$ is obtained by just making a replacement
\begin{equation}
\label{htor}
a^-_j \to -h_{\tau} b_j.
\end{equation}
Note that we cannot do this simple replacement if our expression for $\mathcal{R}$ was not normal-ordered, because the commutation relations satisfied by $a^-_j$ and $b_j$ are different. We can also go back: if we know a normal-ordered expression for $\mathcal{H}$, we can do the same replacement in the other direction to find an expression for the full $\mathcal{R}$. In this sense, the knowledge of normal-ordered expression of $R$-matrix acting on two Fock spaces is the same thing as knowing the normal-ordered expression for the vacuum-to-vacuum matrix element of $R$ in the auxiliary space. For this to work in the simple way described, it is important that the $R$-matrix depends only on the differences of auxiliary and quantum oscillators and commutes with their sum.

We can now diagonalize $\mathcal{H}^{(\tau)}$ in one-boson representation. By our choice of the vacuum vector, we have
\begin{equation}
\mathcal{H}^{(\tau)} \ket{0}_Q = \ket{0}_Q.
\end{equation}
At level one we have a single state $b_{-1} \ket{0}_Q$ on which $\mathcal{H}^{(\tau)}$ acts with an eigenvalue
\begin{equation}
\frac{2a^-_0 \sigma_3 + h_{\tau} h_{\sigma} (h_{\tau}-h_{\sigma})}{2a^-_0 \sigma_3 - h_{\tau} h_{\sigma} (h_{\tau}+h_{\sigma})}.
\end{equation}
At higher levels the eigenstates of $\mathcal{H}^{(\tau)}$ are labeled by partitions since we are considering a single bosonic Fock space. Furthermore, since $\mathcal{H}^{(\tau)}$ mutually commute for any value of the spectral parameter and for all three values of $\tau$, the eigenvectors will only depend on Young diagram label and the choice of $h_\sigma$. Let us for simplicity of notation choose $h_\sigma = h_3$. The first few eigenstates are summarized in the following table:

\begin{center}
\begin{tabular}{|c|c|c|}
\hline
Young d. & state & roots \\
\hline
$(1)$ & $b_{-1}$ & $0$ \\
$(2)$ & $b_{-1}^2 - h_2^{-1} b_{-2}$ & $0, h_1$ \\
$(1,1)$ & $b_{-1}^2 - h_1^{-1} b_{-2}$ & $0, h_2$ \\
$(3)$ & $b_{-1}^3 -3h_2^{-1} b_{-2} b_{-1} + 2h_2^{-2} b_{-3}$ & $0, h_1, 2h_1$ \\
$(2,1)$ & $b_{-1}^3 - (h_1^{-1} + h_2^{-1}) b_{-2} b_{-1} + h_1^{-1} h_2^{-1} b_{-3}$ & $0, h_1, h_2$ \\
$(1,1,1)$ & $b_{-1}^3 - 3h_1^{-1} b_{-2} b_{-1} + 2h_1^{-2} b_{-3}$ & $0, h_2, 2h_2$ \\
$(4)$ & $b_{-1}^4 - 6h_2^{-1} b_{-2} b_{-1}^2 + 3h_2^{-2} b_{-2}^2 + 8h_2^{-2} b_{-3} b_{-1} - 6h_2^{-3} b_{-4}$ & $0, h_1, 2h_1, 3h_1$ \\
$(3,1)$ & $b_{-1}^4 -(h_1^{-1}+3h_2^{-1}) b_{-2} b_{-1}^2 + h_1^{-1} h_2^{-1} b_{-2}^2 + \ldots $ & $0, h_1, 2h_1, h_2$ \\
 & $\ldots + 2h_2^{-1} (h_1^{-1} + h_2^{-1}) b_{-3} b_{-1} - 2h_1^{-1}h_2^{-2} b_{-4}$ & \\
$(2,2)$ & $b_{-1}^4 - 2(h_1^{-1}+h_2^{-1}) b_{-2} b_{-1}^2 + (h_1^{-2} + h_2^{-2} - h_1^{-1} h_2^{-1}) b_{-2}^2 + \ldots$ & $0, h_1, h_2, h_1+h_2$ \\
 & $\ldots + 4h_1^{-1}h_2^{-1} b_{-3} b_{-1} - (h_1^{-1} + h_2^{-1})h_1^{-1} h_2^{-1} b_{-4}$ & \\
$(2,1,1)$ & $b_{-1}^4 -(3h_1^{-1}+h_2^{-1})b_{-2}b_{-1}^2 + h_1^{-1}h_2^{-1} b_{-2}^2 + \ldots$ & $0, h_1, h_2, 2h_2$ \\
 & $\ldots + 2h_1^{-1}(h_1^{-1}+h_2^{-1}) b_{-3} b_{-1} -2h_1^{-2}h_2^{-1} b_{-4}$ & \\
$(1,1,1,1)$ & $b_{-1}^4 - 6h_1^{-1} b_{-2}b_{-1}^2 + 3h_1^{-2} b_{-2}^2 + 8h_1^{-2} b_{-3} b_{-1} - 6h_1^{-3} b_{-4}$ & $0, h_2, 2h_2, 3h_2$ \\
\hline
\end{tabular}
\end{center}
First of all, we notice that the expressions for eigenvectors are closely related to Jack polynomials. It is a known fact that in free field representations of Virasoro algebra one encounters these \cite{mimachi1995singular}, so it thus comes at no surprise to also find them here. For review of properties of Jack polynomials see \cite{macdonald1998symmetric, stanleyjack, Nazarov:2012gn}. What we need here is the fact that Jack polynomials are symmetric polynomials which are a deformation of the well-known Schur polynomials. They form a basis of the vector space of all symmetric functions. Following the conventions of Stanley (i.e. the coefficient of $e_n$ is $n!$), the first few Jack polynomials $J$ expressed as linear combinations of Newton's power sum polynomials $p_j$ are:
\begin{align*}
J_{(1)}^{\alpha} & = p_1 = e_1 \\
J_{(2)}^{\alpha} & = p_1^2 + \alpha p_2 \\
J_{(1,1)}^{\alpha} & = p_1^2 - p_2 = e_2 \\
J_{(3)}^{\alpha} & = 2\alpha^2 p_3 + 3\alpha p_1 p_2 + p_1^3 \\
J_{(2,1)}^{\alpha} & = p_1^3 + (\alpha-1) p_1 p_2 - \alpha p_3 \\
J_{(1,1,1)}^{\alpha} & = p_1^3 - 3p_1 p_2 + 2p_3 = 6e_3 \\
J_{(4)}^{\alpha} & = 6\alpha^3 p_4 + 8\alpha^2 p_1 p_3 + 3\alpha^2 p_2^2 + 6\alpha p_1^2 p_2 + p_1^4 \\
J_{(3,1)}^{\alpha} & = -2\alpha^2 p_4 + 2\alpha(\alpha-1) p_1 p_3 -\alpha p_2^2 + (3\alpha-1) p_1^2 p_2 + p_1^4 \\
J_{(2,2)}^{\alpha} & = \alpha(1-\alpha) p_4 -4\alpha p_1 p_3 + (1+\alpha+\alpha^2) p_2^2 + 2(\alpha-1) p_1^2 p_2 + p_1^4 \\
J_{(2,1,1)}^{\alpha} & = 2\alpha p_4 + 2(1-\alpha) p_1 p_3 - \alpha p_2^2 + (\alpha-3) p_1^2 p_2 + p_1^4 \\
J_{(1^4)}^{\alpha} & = -6 p_4 + 8 p_1 p_3 + 3 p_2^2 -6 p_1^2 p_2 + p_1^4
\end{align*}
The identification between our eigenvectors and Jack polynomials is the following: the deformation parameter $\alpha$ is identified with
\begin{equation}
\label{jackalpha}
\alpha = -\frac{h_1}{h_2}.
\end{equation}
The Newton power sum polynomials are identified with $b_j$ oscillators via
\begin{equation}
\label{jackoscil}
p_j \leftrightarrow h_1 b_{-j}
\end{equation}
and the eigenstates are exactly the Jack polynomials and agree with the normalization of the eigenvectors given in the table if we rescale $J$ by $h_1^{\deg J}$. The Fock representation that we are studying ($Y_{001}$ in the notation of \cite{Gaiotto:2017euk}) has a symmetry under exchange of $h_1 \leftrightarrow h_2$. In terms of Jack polynomials and eigenstates of $\mathcal{H}^{(\tau)}$ this symmetry acts as $\alpha \leftrightarrow \alpha^{-1}$ and is accompanied by a transposition of the Young diagram labeling the corresponding state.

Let's now turn to the spectrum of $\mathcal{H}^{(\tau)}$ operators in this basis. First of all, for each eigenvector the table above gives a list of what we call the roots. They correspond to boxes of the Young diagram and are simply the weighted coordinates of the given box with $h_1$ being the weight associated to the horizontal direction and $h_2$ to the vertical direction. We choose the origin of the coordinates such that the first box (upper left) has weighted coordinate $0$. This type of weighted coordinates of boxes of (plane) partitions appears frequently in the representations of the affine Yangian, see for example \cite{tsymbaliuk2017affine, Prochazka:2015deb}. The eigenvalue of $\mathcal{H}^{(\tau)}$ acting on the eigenstate labeled by Young diagram $\Lambda$ turns out to be
\begin{equation}
\label{htaubosspectrum}
\mathcal{H}^{(\tau)}(u)\ket{\Lambda} = \prod_{\Box \in \Lambda} \frac{u - q - h_{\Box}} {u - q - h_{\Box} + h_{\tau}} \ket{\Lambda}
\end{equation}
where we defined $h_{\Box} = h_1 x_1(\Box) + h_2 x_2(\Box) + h_3 x_3(\Box)$ and
\begin{equation}
\label{uu0}
u_A - q = \frac{h_{\sigma}-h_{\tau}}{2}-\frac{a^-_0 \sigma_3}{h_{\sigma} h_{\tau}} = \frac{h_{\sigma}-h_{\tau}}{2}-\frac{(h_{\sigma} a_{A,0} - h_{\tau} b_0) \sigma_3}{h_{\sigma} h_{\tau}} = u_A +\frac{b_0 \sigma_3}{h_{\sigma}} + \frac{h_{\sigma}}{2}.
\end{equation}
or
\begin{equation}
q = -\frac{b_0 \sigma_3}{h_{\sigma}} - \frac{h_{\sigma}}{2}.
\end{equation}
We see that the spectrum of $\mathcal{H}^{(\tau)}$ is determined combinatorially in terms of the Young diagrams labeling the state. The eigenvalues take a similar product form as the eigenvalues of Yangian $\psi(u)$ generating function, but in the case of $\mathcal{H}^{(\tau)}$ every box in the Young diagram contributes by a single zero and a single pole which are distance $h_{\tau}$ apart. In the case of $\psi(u)$ every box gave contribution $\varphi(u)$ (where $\varphi(u)$ is the structure constant of the Yangian) which had three zeros and three poles, symmetrically shifted in $h_1, h_2$ and $h_3$ directions. It is thus quite easy to express $\psi(u)$ in terms of $\mathcal{H}^{(\tau)}$ which we will do later.

The constant $u_0$ is related to the zero modes of the fields. Shifting a zero mode of free boson by a constant is an automorphism of the free boson Heisenberg algebra. The corresponding symmetry of affine Yangian is a translation of the spectral parameter.5

\subsection{Nazarov-Sklyanin operators}
Having found the eigenvectors of $\mathcal{H}^{(\tau)}$ and identified them with Jack polynomials, it is natural to ask if there are explicit expressions for operators which would act on Jack polynomials with the same spectrum as we found. In fact, such operators have been found in a nice paper by Nazarov and Sklyanin \cite{Nazarov:2012gn} \footnote{These operators have been used in studies of random partitions in \cite{Moll:2015exa}. It would be nice to see if the $\mathcal{R}$-matrix plays any role in that setting.}. The authors modified the differential operators studied by Debiard and Sekiguchi (acting diagonally on Jack polynomials of $n$ variables, $n < \infty$) in such a way that the modified operators are stable as $n \to \infty$. The resulting operators (considered as a generating function depending on a spectral parameter $u$) were then expanded at large values of the spectral parameter. The form of the expansion looks as follows: we have
\begin{align}
\label{nazarova}
\nonumber
A(u_{NS}) & = \, \mathbbm{1} - \frac{1}{u_{NS}} \sum_{j_1>0} M_{j_1} M^\dagger_{j_1} + \frac{1}{2! u_{NS}(u_{NS}+1)} \sum_{j_1,j_2>0} M_{j_1,j_2} M^\dagger_{j_1,j_2} \\
& \, - \frac{1}{3! u_{NS}(u_{NS}+1)(u_{NS}+2)} \sum_{j_1,j_2,j_3>0} M_{j_1,j_2,j_3} M^\dagger_{j_1,j_2,j_3} + \ldots \\
\nonumber
& = \sum_{k=0}^{\infty} \frac{(-1)^k}{k! (u_{NS})_k} \sum_{j_1,\ldots,j_k>0} M_{j_1,\ldots,j_k} M^\dagger_{j_1,\ldots,j_k}
\end{align}
where $M_{j_1,\ldots,m_k}$ are operators of multiplication by symmetric polynomials and $M^\dagger_{j_1,\ldots,m_k}$ are the corresponding annihilation operators acting by differentiation. More precisely, for the first few $M_{j_1,\ldots,m_k}$ we have
\begin{align}
\nonumber
M_{j_1} & = p_{j_1} \\
\nonumber
M_{j_1,j_2} & = p_{j_1} p_{j_2} - p_{j_1+j_2} \\
\nonumber
M_{j_1,j_2,j_3} & = p_{j_1} p_{j_2} p_{j_3} - p_{j_1+j_2} p_{j_3} - p_{j_1+j_3} p_{j_2} - p_{j_2+j_3} p_{j_1} + 2 p_{j_1+j_2+j_3} \\
M_{j_1,j_2,j_3,j_4} & = p_{j_1} p_{j_2} p_{j_3} p_{j_4} - p_{j_1+j_2} p_{j_3} p_{j_4} - p_{j_1+j_3} p_{j_2} p_{j_4} - p_{j_1+j_4} p_{j_2} p_{j_3} - p_{j_2+j_3} p_{j_1} p_{j_4} \\
\nonumber
& \, - p_{j_2+j_4} p_{j_1} p_{j_3} - p_{j_3+j_4} p_{j_1} p_{j_2} + p_{j_1+j_2} p_{j_3+j_4} + p_{j_1+j_3} p_{j_2+j_4} + p_{j_1+j_4} p_{j_2+j_3} \\
\nonumber
& \, + 2 p_{j_1+j_2+j_3} p_{j_4} + 2 p_{j_1+j_2+j_4} p_{j_3} + 2 p_{j_1+j_3+j_4} p_{j_2} + 2 p_{j_2+j_3+j_4} p_{j_1} - 6 p_{j_1+j_2+j_3+j_4}
\end{align}
and the general pattern should be clear: we are summing over all ways of grouping the indices $(j_1,\ldots,j_k)$ with alternating minus signs and with an additional factor of $(l-1)!$ every time $l$ indices were grouped together. If all the indices $j_k$ take different values, this is just the expression of monomial symmetric function in terms of Newton power sum polynomials, but if some of $j_k$'s are equal, there are additional overall combinatorial factors which cancel out in the final formula for $A(u_{NS})$ (see \cite{Nazarov:2012gn} for details on these additional factors). We can write this compactly as
\begin{multline}
M_{j_1,\ldots,j_n} = \Sym_{j_1,\ldots,j_n} \sum_{\sum_{l=1}^k \lambda_l = n} p_{j_1+\ldots+j_{\lambda_1}} p_{j_{\lambda_1+1}+\ldots+j_{\lambda_1+\lambda_2}} \cdots p_{j_{\lambda_1+\ldots+\lambda_{k-1}+1}+\ldots+j_{\lambda_1+\ldots+\lambda_k}} \times \\
\times \frac{(-1)^k n!}{z_\lambda} \times \prod_{l=1}^k (\lambda_l-1)!
\end{multline}
where $z_\lambda$ is the standard combinatorial factor (for partition with $l_j$ cycles of length $j$ it is given by $\prod_j l_j! j^{l_j}$).

The expressions for $M^\dagger_{j_1,\ldots,j_k}$ take formally the same form with $p_j$ replaced by $p^\dagger_j$ where $p^\dagger_j$ are the lowering operators acting as derivatives with respect to $p_j$,
\begin{equation}
p^\dagger_j = \alpha j \frac{\partial}{\partial p_j}.
\end{equation}
We have the obvious commutation relation
\begin{equation}
\left[ p^\dagger_j, p_k \right] = \alpha j \delta_{j,k}.
\end{equation}
so if we identify $p_j$ with $a_{-j}$ and $p^\dagger_k$ with $a_k$ we get exactly the free boson Heisenberg algebra with normalization constant (two point function) $\alpha$. In this sense, (\ref{nazarova}) can be interpreted as normal-ordered expression for an operator acting in the bosonic Fock space. One additional piece of information that Nazarov and Sklyanin provide is the spectrum of $A(u)$ when acting on Jack polynomials. The eigenvalues take the product form
\begin{equation}
\label{ansev}
A(u_{NS}) J_{\lambda} = \prod_{j=1}^n \left( \frac{u_{NS}+j-1-\alpha \lambda_j}{u_{NS}+j-1} \right) J_{\lambda}
\end{equation}
Here $\lambda_j$ are lengths of rows of the Young diagram and $n$ is any sufficiently large integer (it should be at least equal to the number of rows of the Young diagram). It is exactly the denominator factors that were added by Nazarov and Sklyanin to make this expression independent of the number of variables and thus stable in the limit of large number of independent variables.

We want now to identify these operators with $\mathcal{H}^{(\tau)}$ transfer matrices constructed above. First of all, we can rewrite (\ref{ansev}) in the form
\begin{equation}
A(u_{NS}) J_{\lambda} = \prod_{\Box \in \lambda} \left( \frac{-u_{NS} h_2 - h_1 - x_1(\Box) h_1 - x_2(\Box) h_2 - x_3(\Box) h_3}{-u_{NS} h_2 - x_1(\Box) h_1 - x_2(\Box) h_2 - x_3(\Box) h_3} \right) J_{\lambda}
\end{equation}
where we used (\ref{jackalpha}) and introduced formally the third coordinate of the boxes of $\lambda$ in such a way that we can think of the upper left box of $\lambda$ to have coordinates either $(0,0,0)$ or $(1,1,1)$ without changing the result (because $h_1+h_2+h_3=0$). Comparing this to the spectrum of (\ref{htaubosspectrum}) we see that we need to put $h_{\tau} = h_1$ and still $h_{\sigma} = h_3$ and furthermore
\begin{equation}
u_A-q = -u_{NS}h_2 - h_1
\end{equation}
or
\begin{equation}
u_{NS} = a^-_0 + \frac{1}{2}.
\end{equation}
With these identifications and the identification
\begin{equation}
h_1 b_{-k} \leftrightarrow p_k, \quad\quad\quad -h_2 b_k \leftrightarrow k \frac{\partial}{\partial p_k}, \quad\quad\quad h_1 b_k \leftrightarrow p^\dagger_k
\end{equation}
we can identify the operator $\mathcal{H}^{(1)}$ with Nazarov-Sklyanin operator $A(u_{NS})$,
\begin{equation}
\mathcal{H}^{(1)}(a^-_0) = A(a^-_0+1/2).
\end{equation}
The expansion of $\mathcal{H}^{(1)}$ is thus
\begin{align}
\label{h1expansionns}
\mathcal{H}^{(1)}(a^-_0) & = \mathbbm{1} - \frac{1}{(a^-_0+1/2)} \sum_{j_1>0} h_1^2 b_{-j_1} b_{j_1} \\
\nonumber
& + \frac{1}{2!\,(a^-_0+1/2)(a^-_0+3/2)} \sum_{j_1,j_2>0} (h_1^2 b_{-j_1} b_{-j_2} - h_1 b_{-j_1-j_2}) (h_1^2 b_{j_1} b_{j_2} - h_1 b_{j_1+j_2}) + \ldots
\end{align}
Finally, we may use the replacement rule (\ref{htor}) to reconstruct the whole $\mathcal{R}$-matrix (of the mixed type),
\begin{align}
\label{rmix}
\nonumber
\mathcal{R} & = \mathbbm{1} - \frac{1}{(a^-_0+1/2)} \sum_{j_1>0} a^-_{-j_1} a^-_{j_1} \\
& + \frac{1}{2!\,(a^-_0+1/2)(a^-_0+3/2)} \sum_{j_1,j_2>0} (a^-_{-j_1} a^-_{-j_2} + a^-_{-j_1-j_2}) (a^-_{j_1} a^-_{j_2} + a^-_{j_1+j_2}) \\
\nonumber
& - \frac{1}{3!\,(a^-_0+1/2)(a^-_0+3/2)(a^-_0+5/2)} \sum_{j_1,j_2,j_3>0} \mathcal{M}_{-j_1,-j_2,-j_3} \mathcal{M}_{j_1,j_2,j_3} + \ldots
\end{align}
where we used the same shorthand notation as in (\ref{rmixexp}). We should compare this to the results of section \ref{secRuexp}, in particular the equation (\ref{rmixexp}) and indeed expanding (\ref{rmix}) at large $a_0$ we reproduced exactly all the terms in (\ref{rmixexp}).

To summarize, we have used the known eigenstates of the transfer matrix obtained from the mixed $\mathcal{R}$-matrix and the Nazarov-Sklyanin formula for the generating function of commuting Hamiltonians acting on Jack polynomials to find a closed-form expression for the mixed $\mathcal{R}$-matrix, equation (\ref{rmix}). This is one of the most important results of this article.

\subsection{Ladder operators}
We can now turn to ladder operators of $\mathcal{W}_{1+\infty}$ and relate them to the generators of the Yangian algebra associated to Fock $\mathcal{R}$-matrix.

\paragraph{Ladder operators in the Yangian presentation}
First of all, we know that in Arbesfeld-Schiffmann-Tsymbaliuk presentation of $\mathcal{W}_{1+\infty}$ there are ladder operators $e(u)$ and $f(u)$ such that in representations on plane partitions these act as \cite{arbesschif,tsymbaliuk2017affine,Prochazka:2015deb}
\begin{align}
\label{effockaction}
e(u) \ket{\Lambda} & = \sum_{\Box \in \Lambda^+} \frac{E(\Lambda\to\Lambda+\Box)}{u-q-h_{\Box}} \ket{\Lambda+\Box} \\
f(u) \ket{\Lambda} & = \sum_{\Box \in \Lambda^-} \frac{F(\Lambda\to\Lambda-\Box)}{u-q-h_{\Box}} \ket{\Lambda-\Box},
\end{align}
i.e. the action of $e(u)$ on a state labeled by a certain plane partition $\Lambda$ gives a linear combination of states labeled by all possible plane partitions obtained from $\Lambda$ by adding a box (in way compatible with plane partition rules). $E(\Lambda\to\Lambda+\Box)$ are certain probability amplitudes of this process and depend on the choice of the normalization of vectors $\ket{\Lambda}$. On the other hand, the spectral parameter dependence of these formulas is completely fixed by the relations of $\mathcal{W}_{1+\infty}$.

In the case at hand (representation of $\mathcal{W}_{1+\infty}$ on one free boson) we can even find the amplitudes $E$ and $F$. First of all, we can simplify the calculation by putting $q=0$ (we can always reconstruct the general case by shifting the spectral parameter). After this is done, the amplitudes $E$ can be fixed from (\ref{effockaction}) by only looking at the leading coefficient of $e(u)$ as $u\to\infty$, i.e.
\begin{equation}
e_0 \ket{\Lambda} = \sum_{\Box \in \Lambda^+} E(\Lambda\to\Lambda+\Box) \ket{\Lambda+\Box}.
\end{equation}
Choosing for concreteness $\ket{\Lambda}$ to be the Jack polynomials normalized as in the table above and identifying \footnote{This choice of normalization is slightly different than the one used in \cite{Prochazka:2015deb}, but it is equivalent to this one after applying the scaling symmetry of the algebra.}
\begin{equation}
e_0 = b_{-1}, \quad\quad\quad f_0 = -b_1,
\end{equation}
we find for the action
\begin{align}
\label{ffe0norm}
e_0 \ket{\Lambda} = \sum_{\Box\in\Lambda^+} \frac{1}{h_{\Box}} \res_{u=h_{\Box}} \left( \prod_{\Box^\prime \in \Lambda} \frac{(u-h_{\Box^\prime}) (u-h_{\Box^\prime}-h_1-h_2)}{(u-h_{\Box^\prime}-h_1) (u-h_{\Box^\prime}-h_2)} \right) \ket{\Lambda+\Box}
\end{align}
Few concrete examples of application of this formula are given in appendix \ref{appfock}. Analogously, for the annihilation amplitudes we find
\begin{align}
\label{fff0norm}
f_0 \ket{\Lambda} = \sum_{\Box\in\Lambda^-} \frac{h_{\Box}+h_1+h_2}{h_1^2 h_2^2} \res_{u=h_{\Box}} \left( \prod_{\Box^\prime \in \Lambda} \frac{(u-h_{\Box^\prime}+h_1) (u-h_{\Box^\prime}+h_2)}{(u-h_{\Box^\prime}) (u-h_{\Box^\prime}+h_1+h_2)} \right) \ket{\Lambda-\Box}.
\end{align}

\paragraph{Ladder operators from the YBE}
Let us now take a matrix element of equation (\ref{ybehe}) between the states $\bra{\Lambda+\Box}$ and $\ket{\Lambda}$. Denoting the matrix element of $\mathcal{E}^{(\tau)}$ by the same symbol, we find using (\ref{htaubosspectrum}) the condition
\begin{multline}
\frac{(u-v+h_\tau)(u-q-h_\Box)}{(u-q-h_\Box+h_\tau)} \prod_{\Box^\prime \in \Lambda} \left( \frac{u-q-h_{\Box^\prime}}{u-q-h_{\Box^\prime}+h_\tau} \right) \mathcal{E}^{(\tau)}(v) = \\ (u-v) \prod_{\Box^\prime \in \Lambda} \left( \frac{u-q-h_{\Box^\prime}}{u-q-h_{\Box^\prime}+h_\tau} \right) \mathcal{E}^{(\tau)}(v) + \frac{h_\tau(v-q-h_\Box)}{(v-q-h_\Box+h_\tau)} \prod_{\Box^\prime \in \Lambda} \left( \frac{v-q-h_{\Box^\prime}}{v-q-h_{\Box^\prime}+h_\tau} \right) \mathcal{E}^{(\tau)}(u)
\end{multline}
which determines $\mathcal{E}^{(\tau)}(\Lambda\to\Lambda+\Box)$ uniquely up to a $u$-independent factor to be
\begin{align}
\label{ffecalact}
\mathcal{E}^{(\tau)}(\Lambda\to\Lambda+\Box)(u) & \sim \frac{1}{u-q-h_{\Box}+h_{\tau}} \prod_{\Box^\prime \in \Lambda} \frac{u-q-h_{\Box^\prime}}{u-q-h_{\Box^\prime}+h_{\tau}} \\
\nonumber
& \sim \frac{1}{u-q-h_{\Box}} \prod_{\Box^\prime \in \Lambda+\Box} \frac{u-q-h_{\Box^\prime}}{u-q-h_{\Box^\prime}+h_{\tau}}
\end{align}
with $q$ as in (\ref{uu0}) and $u \equiv u_A$.

We can compare this to the large $u$ expansion of $\mathcal{E}^{(\tau)}$ obtained from the $\mathcal{R}$-matrix. For the first few terms, we find
\begin{align}
\nonumber
\mathcal{E}^{(\tau)} = & -\frac{h_{\tau} h_{\sigma}}{\sigma_3 a^-_0} h_{\tau} b_{-1} - \frac{1}{(a_0^-)^2} \frac{h_{\tau}h_{\sigma}}{\sigma_3} \left[ \sum_{j>0} (h_{\tau}^2 b_{-j-1} -h_{\tau}^3 b_{-j} b_{-1}) b_j - \frac{\rho}{2} h_\tau b_{-1} \right] + \\
\nonumber
& + \frac{h_{\tau} h_{\sigma}}{\sigma_3 (a^-_0)^3} \Big[ \frac{1}{2} \sum_{j,k>0} (-h_\tau^3 b_{-j} b_{-k} b_{-1} + h_\tau^2 b_{-j-k} b_{-1} + h_\tau^2 b_{-j-1} b_{-k} \\
& + h_\tau^2 b_{-k-1} b_{-j} - 2h_\tau b_{-j-k-1}) \times (h_\tau^2 b_j b_k - h_\tau b_{j+k}) \\
\nonumber
& -2 h_\tau^3 \sum_{j>0} b_{-1} b_{-j} b_j - h_\tau^3 \frac{\rho-1}{2} \sum_{j>0} (j+1) b_{-1} b_{-j} b_j + 2h_\tau^2 \sum_{j>0} b_{-j-1} b_j \\
\nonumber
& + (\rho-1) h_\tau^2 \sum_{j>0} (j+1) b_{-j-1} b_j - \frac{h_\tau}{4} b_{-1} +h_\tau (\rho-1)(1-3\rho-4)/12 b_{-1} \Big] + \cdots
\end{align}
For lower lying states it is easy to check that this expansion exactly acts as in (\ref{ffecalact}). We can use the action (\ref{ffe0norm}) together with the fact that the leading term in the large $u$ expansion of $\mathcal{E}(u)$ is
\begin{equation}
\mathcal{E}^{(\tau)}(u) \sim \frac{h_\tau e_0}{u} + \mathcal{O}\left(\frac{1}{u^2}\right)
\end{equation}
to fix the normalization. We find
\begin{multline}
\mathcal{E}^{(\tau)}(u) \ket{\Lambda} = \sum_{\Box\in\Lambda^+} \frac{h_\tau}{h_{\Box}} \frac{1}{u-q-h_{\Box}+h_{\tau}} \prod_{\Box^\prime \in \Lambda} \left( \frac{u-q-h_{\Box^\prime}}{u-q-h_{\Box^\prime}+h_{\tau}} \right) \times \\
\times \res_{u=h_{\Box}} \left( \prod_{\Box^\prime \in \Lambda} \frac{(u-h_{\Box^\prime}) (u-h_{\Box^\prime}-h_1-h_2)}{(u-h_{\Box^\prime}-h_1) (u-h_{\Box^\prime}-h_2)} \right) \ket{\Lambda+\Box}
\end{multline}
where we still have $h_\sigma = h_3$. Comparing this result with (\ref{htaubosspectrum}) and (\ref{effockaction}), we find an operator relation between Arbesfeld-Schiffmann-Tsymbaliuk generating function $e(u)$ and our Yangian generator $\mathcal{E}(u)$,
\begin{equation}
\label{EHtoe1}
e(u) = h_\tau^{-1} \mathcal{E}^{(\tau)}(u-h_\tau) \left( \mathcal{H}^{(\tau)}(u-h_\tau) \right)^{-1}.
\end{equation}
Note in particular that the left-hand side does not depend on our choice of $\tau$. We could have also multiplied $\mathcal{E}$ and $\mathcal{H}$ in the opposite order, finding
\begin{equation}
\label{EHtoe2}
e(u) = h_\tau^{-1} \left( \mathcal{H}^{(\tau)}(u) \right)^{-1} \mathcal{E}^{(\tau)}(u).
\end{equation}
Analogously, the functional equation (\ref{ybehf}) implies that
\begin{align}
\mathcal{F}^{(\tau)}(\Lambda+\Box\to\Lambda)(u) & \sim \frac{1}{u-q-h_{\Box}+h_{\tau}} \prod_{\Box^\prime \in \Lambda} \frac{u-q-h_{\Box^\prime}}{u-q-h_{\Box^\prime}+h_{\tau}} \\
& \sim \frac{1}{u-q-h_{\Box}} \prod_{\Box^\prime \in \Lambda+\Box} \frac{u-q-h_{\Box^\prime}}{u-q-h_{\Box^\prime}+h_{\tau}}.
\end{align}
At large $u$ we have the expansion
\begin{align}
\nonumber
\mathcal{F}^{(\tau)}(u) & = -\frac{h_\tau h_\sigma}{\sigma_3 a^-_0} h_\tau b_1 - \frac{h_\tau h_\sigma}{\sigma_3 (a^-_0)^2} \left[ \sum_{j>0} (h_\tau^2 b_{-j} b_{j+1} - h_\tau^3 b_{-j} b_j b_1) - \frac{\rho}{2} h_\tau b_1 \right] + \\
\nonumber
& + \frac{h_{\tau} h_{\sigma}}{\sigma_3 (a^-_0)^3} \Big[ \frac{1}{2} \sum_{j,k>0} (h_\tau^2 b_{-j} b_{-k} - h_\tau b_{-j-k}) \times \\
& \times (-h_\tau^3 b_j b_k b_1 + h_\tau^2 b_{j+k} b_1 + h_\tau^2 b_{j+1} b_k + h_\tau^2 b_{k+1} b_j - 2h_\tau b_{j+k+1}) \\
\nonumber
& -2 h_\tau^3 \sum_{j>0} b_{-j} b_j b_1 - h_\tau^3 \frac{\rho-1}{2} \sum_{j>0} (j+1) b_{-j} b_1 b_j + 2h_\tau^2 \sum_{j>0} b_{-j} b_{j+1} \\
\nonumber
& + (\rho-1) h_\tau^2 \sum_{j>0} (j+1) b_{-j} b_{j+1} - \frac{h_\tau}{4} b_1 +h_\tau (\rho-1)(1-3\rho-4)/12 b_1 \Big] + \cdots.
\end{align}
By comparing the leading order coefficient with (\ref{fff0norm})
\begin{multline}
\mathcal{F}^{(\tau)}(u) \ket{\Lambda} = \sum_{\Box\in\Lambda^-} \frac{h_{\Box}+h_1+h_2}{h_1^2 h_2^2} \frac{-h_\tau}{u-q-h_\Box+h_\tau} \left(\prod_{\Box^\prime \in \Lambda-\Box} \frac{u-q-h_{\Box^\prime}}{u-q-h_{\Box^\prime}+h_\tau}\right) \times \\
\times \res_{u=h_{\Box}} \left( \prod_{\Box^\prime \in \Lambda} \frac{(u-h_{\Box^\prime}+h_1) (u-h_{\Box^\prime}+h_2)}{(u-h_{\Box^\prime}) (u-h_{\Box^\prime}+h_1+h_2)} \right) \ket{\Lambda-\Box}.
\end{multline}
The corresponding Arbesfeld-Schiffmann-Tsymbaliuk generating function $f(u)$ is given by
\begin{align}
\nonumber
\label{FHtof}
f(u) & = -h_\tau^{-1} \left( \mathcal{H}^{(\tau)}(u-h_\tau) \right)^{-1} \mathcal{F}^{(\tau)}(u-h_\tau) \\
& = -h_\tau^{-1} \mathcal{F}^{(\tau)}(u) \left( \mathcal{H}^{(\tau)}(u) \right)^{-1}.
\end{align}
We have thus identified the Arbesfeld-Schiffmann-Tsymbaliuk generating functions in terms of the Yangian generators (matrix elements of the Fock $\mathcal{R}$-matrix).

\subsection{Nazarov-Sklyanin II}
Before closing this section, let us mention another interesting operator constructed by Nazarov and Sklyanin \cite{nazarovskl}. Consider an infinite matrix with non-commuting matrix elements
\begin{equation}
\label{nsdeterminant}
\mathcal{L}(u) = \begin{pmatrix}
0 & \gamma b_{-1} & \gamma b_{-2} & \gamma b_{-3} & \gamma b_{-4} & \ldots \\
\gamma b_1 & h_1+h_2 & \gamma b_{-1} & \gamma b_{-2} & \gamma b_{-3} & \ldots \\
\gamma b_2 & \gamma b_1 & 2(h_1+h_2) & \gamma b_{-1} & \gamma b_{-2} & \ldots \\
\gamma b_3 & \gamma b_2 & \gamma b_1 & 3(h_1+h_2) & \gamma b_{-1} & \ldots \\
\ldots & \ldots & \ldots & \ldots & \dots & \ldots \\
\end{pmatrix}
\end{equation}
with $\gamma = -h_1 h_2$ and take the the upper left matrix element of its resolvent
\begin{equation}
\mathcal{U}(u) = \left(\frac{u}{u-\mathcal{L}}\right)_{00} \equiv 1 + u^{-1} \mathcal{L}_{00} + u^{-2} \sum_{j \geq 0} \mathcal{L}_{0j} \mathcal{L}_{j0} + u^{-3} \sum_{j,k \geq 0} \mathcal{L}_{0j} \mathcal{L}_{jk} \mathcal{L}_{k0} + \ldots
\end{equation}
considered as Laurent expansion at infinite spectral parameter $u$. More concretely
\begin{align}
\nonumber
\mathcal{U}(u) & = 1 + u^{-2} h_1^2 h_2^2 \sum_{j>0} b_{-j} b_j \\
\nonumber
& + u^{-3} \Big( -h_1^3 h_2^3 \sum_{j,k>0} (b_{-j} b_{-k} b_{j+k} + b_{-j-k} b_j b_k) + h_1^2 h_2^2 (h_1+h_2) \sum_{j>0} j b_{-j} b_j \Big) \\
\nonumber
& + u^{-4} \Big( h_1^4 h_2^4 \sum_{j,k,l>0} (b_{-j-k-l} b_j b_k b_l + b_{-j} b_{-k} b_{-l} b_{j+k+l} + 2 b_{-j} b_{-k-l} b_{j+k} b_l \\
& + b_{-j-k} b_k b_{-k-l} b_{j+k+l} + b_{-j-k-l} b_{k+l} b_{-k} b_{j+k}) \\
\nonumber
& + h_1^4 h_2^4 \sum_{j,k>0} (b_{-j} b_j b_{-j-k} b_{j+k} + b_{-j-k} b_{j+k} b_{-j} b_j + b_{-j} b_{-k} b_j b_k + b_{-j-k} b_k b_{-k} b_{j+k}) \\
\nonumber
& - \frac{3}{2} h_1^3 h_2^3 (h_1+h_2) \sum_{j,k>0} (j+k)(b_{-j-k} b_j b_k + b_{-j} b_{-k} b_{j+k}) + h_1^4 h_2^4 \sum_{j>0} b_{-j} b_j b_{-j} b_j \\
\nonumber
& + h_1^2 h_2^2 (h_1+h_2)^2 \sum_{j>0} j^2 b_{-j} b_j \Big) + \mathcal{O}\left(u^{-5}\right)
\end{align}
It was shown in \cite{nazarovskl} that the action of $\mathcal{U}(u)$ on Jack polynomials analogously to (\ref{htaubosspectrum}) is diagonal with eigenvalues
\begin{equation}
\mathcal{U}(u)\ket{\Lambda} = \prod_{\Box \in \Lambda} \frac{(u - h_{\Box})(u - h_{\Box} - h_1 - h_2)}{(u - h_{\Box} - h_1)(u - h_{\Box} - h_2)} \ket{\Lambda}.
\end{equation}
These operators give us another interesting family of commuting operators acting diagonally on Jack polynomials.

\section{$\mathcal{R}$-matrix from fermions}
\label{secfermion}
We can derive another expression for the mixed $R$-matrix starting from the complex fermion. We start with the observation that at $h_1 = 1 = -h_2$ and $h_3 = 0$ the Jack polynomials simplify to Schur polynomials. Furthermore, these are mapped very simply under boson-fermion correspondence to states in the fermionic Fock space:
\begin{center}
\begin{tabular}{|c|c|c|}
\hline
Young diagram & Schur polynomial & fermionic state \\
\hline
$(1)$ & $b_{-1}$ & $\bar{\Psi}_{-1/2} \Psi_{-1/2}$ \\
$(2)$ & $\frac{1}{2}(b_{-1}^2+b_{-2})$ & $\bar{\Psi}_{-3/2} \Psi_{-1/2}$ \\
$(1,1)$ & $\frac{1}{2}(b_{-1}^2-b_{-2})$ & $-\bar{\Psi}_{-1/2} \Psi_{-3/2}$ \\
$(3)$ & $\frac{1}{6}(b_{-1}^3+3b_{-1}b_{-2}+2b_{-3})$ & $\bar{\Psi}_{-5/2} \Psi_{-1/2}$ \\
$(2,1)$ & $\frac{1}{3}(b_{-1}^3-b_{-3})$ & $-\bar{\Psi}_{-3/2} \Psi_{-3/2}$ \\
$(1,1,1)$ & $\frac{1}{6}(b_{-1}^3-3b_{-1}b_{-2}+2b_{-3})$ & $\bar{\Psi}_{-1/2} \Psi_{-5/2}$ \\
$(4)$ & $\frac{1}{24}(b_{-1}^4+6b_{-1}^2b_{-2}+3b_{-2}^2+8b_{-1}b_{-3}+6b_{-4})$ & $\bar{\Psi}_{-7/2} \Psi_{-1/2}$ \\
$(3,1)$ & $\frac{1}{8}(b_{-1}^4+2b_{-1}^2b_{-2}-b_{-2}^2-2b_{-4})$ & $-\bar{\Psi}_{-5/2} \Psi_{-3/2}$ \\
$(2,2)$ & $\frac{1}{12}(b_{-1}^4+3b_{-2}^2-4b_{-1}b_{-3})$ & $\bar{\Psi}_{-3/2} \bar{\Psi}_{-1/2} \Psi_{-3/2} \Psi_{-1/2}$ \\
$(2,1,1)$ & $\frac{1}{8}(b_{-1}^4-2b_{-1}^2b_{-2}-b_{-2}^2+2b_{-4})$ & $\bar{\Psi}_{-3/2} \Psi_{-5/2}$ \\
$(1,1,1,1)$ & $\frac{1}{24}(b_{-1}^4-6b_{-1}^2b_{-2}+3b_{-2}^2+8b_{-1}b_{-3}-6b_{-4})$ & $-\bar{\Psi}_{-1/2} \Psi_{-7/2}$ \\
\hline
\end{tabular}
\end{center}
Here we use the fact that for a pair of complex fermions with OPE
\begin{equation}
\overline{\Psi}(z) \Psi(w) \sim \frac{1}{z-w}
\end{equation}
the corresponding current operator
\begin{equation}
J(z) = (\overline{\Psi} \Psi)(z)
\end{equation}
satisfies the same OPE as the free boson current $J(z) = i\partial\phi(z)$ normalized as
\begin{equation}
J(z) J(w) \sim \frac{1}{(z-w)^2}.
\end{equation}
In terms of mode operators we have
\begin{equation}
b_m = \sum_{k \in \mathbbm{Z}+1/2} : \overline{\Psi}_{m-k} \Psi_k :.
\end{equation}
Using this explicit expression for the current mode operators, we can verify the expressions given in the table. The right-hand side can be obtained also directly from the associated Young diagrams: for that we use the Frobenius notation for the Young diagrams, counting the arm and leg lengths of the diagonal boxes of the Young diagram. In the fermionic Fock space the state associated to Young diagram with Frobenius coordinates $(a_1,a_2,\ldots,a_r;l_1,l_2,\ldots,l_r)$ corresponds up to a sign to a state \cite{miwa2000solitons,Alexandrov:2012tr,Zabrodin:2018uwz}
\begin{equation}
\overline{\Psi}_{-a_r-1/2} \cdots \overline{\Psi}_{-a_1-1/2} \Psi_{-l_r-1/2} \cdots \Psi_{-l_1-1/2} \ket{0}.
\end{equation}
Consider now a generating function
\begin{equation}
\mathcal{H}^{(1)}(u) = \prod_{m \geq \frac{1}{2}} \frac{\left(1 - u^{-1}(m-\frac{1}{2})\right)^{\overline{\Psi}_{-m} \Psi_m}}{\left(1 + u^{-1}(m+\frac{1}{2})\right)^{\Psi_{-m} \overline{\Psi}_m}}.
\end{equation}
We don't need to worry about the ordering on the right-hand side because all the operators in exponents commute. Also, for any fermionic Fock state with bounded Virasoro level there is only a finite number of terms which don't act as identity so we don't need to worry about the convergence. These operators act diagonally on the fermionic representatives of Schur polynomials, the eigenvalues being given by
\begin{equation}
\prod_{\Box \in \Lambda} \frac{u-h_{\Box}}{u-h_{\Box}+h_1}
\end{equation}
which is of the form (\ref{htaubosspectrum}). Note that we still have $h_1 = 1 = -h_2$ and $h_3 = 0$ and for now (for simplicity) we restrict to subspace of states which have $q=0$.

We want to bosonize this expression and write it in the normal-ordered form so that we can deform the expression to the case $h_3 \neq 0$. First of all, we write
\begin{equation}
\mathcal{H}^{(1)}(u) = \exp \left( -\sum_{m \in \mathbbm{Z}+1/2} :\Psi_{-m} \bar{\Psi}_m: \log \left[1+u^{-1}\left(m+\frac{1}{2}\right)\right
] \right).
\end{equation}
Using the mode expansion of $\Psi(z)$ in the complex plane,
\begin{equation}
\bar{\Psi}(z) = \sum_{m \in \mathbbm{Z}+1/2} \bar{\Psi}_m z^{-m-1/2}
\end{equation}
we can write this as
\begin{equation}
\mathcal{H}^{(1)}(u) = \exp \left[ - \oint_0 \frac{dw}{2\pi i} \oint_w \frac{dz}{2\pi i} \frac{1}{z-w} \Psi(z) \log \left( 1 - u^{-1} w \partial_w \right) \bar{\Psi}(w) \right].
\end{equation}
This operator can now be bosonized using the identification \cite{miwa2000solitons,Alexandrov:2012tr,Zabrodin:2018uwz}
\begin{equation}
\overline{\Psi}(z) = : e^{i\phi(z)} : \quad\quad\quad \Psi(z) = :e^{-i\phi(z)}:.
\end{equation}
We find
\begin{equation}
\mathcal{H}^{(1)}(u) = \exp \left[ - \oint_0 \frac{dw}{2\pi i} \oint_w \frac{dz}{2\pi i} \frac{1}{z-w} :e^{-i\phi(z)}: \log \left( 1 - u^{-1} w \partial_w \right) :e^{i\phi(w)}: \right].
\end{equation}
We can integrate by parts in the $w$ integral, introducing the kernel
\begin{equation}
\kappa(u;z,w) = \log \left( 1 + u^{-1} \partial_w w \right) \frac{1}{z-w} = \frac{z}{u(z-w)^2} - \frac{z(z+w)}{2u^2(z-w)^3} + \frac{z(z^2+4zw+w^2)}{3u^3(z-w)^4} + \ldots
\end{equation}
Note that the expansion is not necessarily convergent and we will treat it as a formal power series at $u = \infty$ with coefficients rational functions of $z$ and $w$ singular only on the diagonal $z=w$. Using this kernel, we write
\begin{equation}
\mathcal{H}^{(1)}(u) = \exp \left[ - \oint_0 \frac{dw}{2\pi i} \oint_w \frac{dz}{2\pi i} \kappa(u;z,w) :e^{-i\phi(z)}: :e^{i\phi(w)}: \right].
\end{equation}
In order to write it in explicitly normal ordered form, we first Taylor expand the exponential,
\begin{equation}
\mathcal{H}^{(1)}(u) = \sum_{n=0}^\infty \frac{(-1)^n}{n!} \prod_{j=1}^n \left[ \oint_0 \frac{dw_j}{2\pi i} \oint_{w_j} \frac{dz_j}{2\pi i} \kappa(u;z_j,w_j) :e^{-i\phi(z_j)}: :e^{i\phi(w_j)}: \right].
\end{equation}
We can now use the Wick theorem
\begin{equation}
:e^{i\alpha\phi(z)}: :e^{i\beta\phi(w)}: = (z-w)^{\alpha\beta} :e^{i\alpha\phi(z)} e^{i\beta\phi(w)}:
\end{equation}
to write
\begin{align}
\nonumber
\label{h1fromfermion}
\mathcal{H}^{(1)}(u) = & \sum_{n=0}^\infty \frac{(-1)^{\frac{n(n+1)}{2}}}{n!} \prod_{j=1}^n \left[ \oint_0 \frac{dw_j}{2\pi i} \oint_{w_j} \frac{dz_j}{2\pi i} \kappa(u;z_j,w_j) \right] \times \\
& \times \frac{\prod_{j<k} (z_j-z_k) (w_j-w_k)}{\prod_{j,k} (z_j-w_k)} : e^{-i\sum_j \phi(z_j) + i \sum_j \phi(w_j)}:.
\end{align}
Our convention for $w_j$ integrals around the origin is such that $|w_n|>|w_{n-1}|>\ldots>|w_2|>|w_1|$. This is important when taking the residues to evaluate the terms in $1/u$ expansion of the expression. After evaluating the $z_j$-integrals by taking the residues, we must take the $w_j$-residues in a correct order (the order of taking residues matters because of the diagonal singularities at $w_j = w_k$). We also have to be careful about the extra $(-1)^{\frac{n(n-1)}{2}}$ sign coming from the application of the Wick theorem if we order the variables in Vandermonde determinant as we do above. The expression obtained so far is very reminiscent of the expressions for solitons of KP or 2d Toda hierarchy in terms of tau functions \cite{miwa2000solitons,Alexandrov:2012tr,Zabrodin:2018uwz} which should not be surprising because the tau functions are matrix elements of exponential of fermion bilinears ($GL(\infty)$ group element) and our initial expression for $\mathcal{H}^{(1)}(u)$ was exactly of this form.

We can verify the correctness of (\ref{h1fromfermion}) by expanding it at large $u$. Using equations like
\begin{equation}
\res_{w_2 = 0} \left(\res_{w_1 = 0} \frac{w_1^2-w_1 w_2+w_2^2}{2(w_1-w_2)^2} :J(w_1)J(w_2): \right) = \frac{1}{2} \sum_{j>0} j b_{-j} b_j
\end{equation}
(putting the zero modes $b_0$ to zero) we find
\begin{align}
\nonumber
\mathcal{H}^{(1)}(u) & = 1 + u^{-1} \Big[ -\sum_{j>0} b_{-j} b_j \Big] + u^{-2} \Big[ - \frac{1}{2} \sum_{j,k>0} (b_{-j} b_{-k} b_{j+k} + b_{-j-k} b_j b_k) \\
\nonumber
& + \frac{1}{2} \sum_{j>0} (j+1) b_{-j} b_j + \frac{1}{2} \sum_{j,k>0} b_{-j} b_{-k} b_j b_k \Big] \\
\nonumber
& + u^{-3} \Big[ \sum_{j>0} (-\frac{1}{6} - \frac{1}{2}j -\frac{1}{3}j^2) b_{-j} b_j + \frac{1}{2} \sum_{j,k>0} (j+k+1) (b_{-j} b_{-k} b_{j+k} + b_{-j-k} b_j b_k) \\
& - \frac{1}{3} \sum_{j,k,l>0} (b_{-j} b_{-k} b_{-l} b_{j+k+l} + b_{-j-k-l} b_j b_k b_l) - \frac{1}{2} \sum_{j+k=l+m} b_{-j} b_{-k} b_l b_m \\
\nonumber
& - \frac{1}{4} \sum_{j,k>0} (j+k+2) b_{-j} b_{-k} b_j b_k + \frac{1}{2} \sum_{j,k,l>0} b_{-j} b_{-k} b_{-l} b_{j+k} b_l \\
\nonumber
& - \frac{1}{6} \sum_{j,k,l>0} b_{-j} b_{-k} b_{-l} b_j b_k b_l \Big] + \ldots
\end{align}
or
\begin{align}
\nonumber
\mathcal{H}^{(1)}(u) & = 1 - \frac{1}{u+1} \sum_{j>0} b_{-j} b_j + \frac{1}{2!(u+1)(u+2)} \sum_{j,k>0} (b_{-j} b_{-k} - b_{-j-k})(b_j b_k - b_{j+k}) \\
\nonumber
& - \frac{1}{3!(u+1)(u+2)(u+3)} \sum_{j,k,l>0} (b_{-j} b_{-k} b_{-l} - b_{-j-k} b_{-l} - b_{-j-l} b_{-k} - b_{-k-l} b_{-j} + 2b_{-j-k-l}) \times \\
& \times (b_j b_k b_l - b_{j+k} b_l - b_{j+l} b_k - b_{k+l} b_j + 2b_{j+k+l}) + \ldots
\end{align}
which is exactly coincides with the expansion (\ref{h1expansionns}) with $h_\sigma = h_3 = 0$, $q=0$ and $h_\tau = h_1 = 1$. In this case $a_0^- = u+1/2$ and the formula nicely agrees with the previous result.

\paragraph{Charged states}
Until now we focused on a subsector of the fermionic Fock space which had total charge zero. But we can also consider other subsectors with $b_0 \in \mathbbm{Z}$. The charged vacua are obtained by acting on an uncharged vacuum $\ket{0}$ by fermions of the same charge
\begin{equation}
\ket{n} \sim \overline{\Psi}_{-n+1/2} \cdots \overline{\Psi}_{-3/2} \overline{\Psi}_{-1/2} \ket{0}, \quad\quad\quad n>0
\end{equation}
with charge $q=n$ and
\begin{equation}
\ket{-n} \sim \Psi_{-n+1/2} \cdots \Psi_{-3/2} \Psi_{-1/2} \ket{0}, \quad\quad\quad n>0.
\end{equation}
of charge $q=-n$. The excited states are then obtained by acting by the modes of the current $b_j$ and are still labeled by Young diagrams. The operator $\mathcal{H}^{(1)}(u)$ now acts in the charged Fock space with eigenvalues
\begin{equation}
\label{chargedbosonspectrum}
\frac{\Gamma(u+1)}{u^q \Gamma(u-q+1)} \prod_{\Box \in \Lambda} \frac{u-q-h_{\Box}}{u-q-h_{\Box}+h_1}.
\end{equation}
The prefactor is the $\mathcal{H}^{(1)}$ eigenvalue of the corresponding charged vacuum. It has a well-defined large $u$ expansion
\begin{equation}
\frac{\Gamma(u+1)}{u^q \Gamma(u-q+1)} \sim 1 - u^{-1} {q \choose 2} + u^{-2} {q \choose 3} \frac{3q-1}{4} - u^{-3} {q \choose 4} \frac{q(q-1)}{2} + \ldots.
\end{equation}
With our choice of $h_1 = 1 = -h_2$ and $h_3 = 0$ we have a simple identification $b_0 = q$ and it is straightforward to verify that (\ref{h1fromfermion}) acts correctly also on the charged bosonic Fock space (\ref{chargedbosonspectrum}).

\paragraph{Mixed $\mathcal{R}$-matrix}
We can now use the fact that the normal-ordered large $u$ expansions of the Nazarov-Sklyanin operators (\ref{h1expansionns}) for all values of $h_1$ (not just $h_1=1$ as we assumed until now) are trivially related by just rescaling the current. This means that we only need to make the replacements
\begin{align}
\nonumber
\label{fermrrepl}
\phi(z) & \rightarrow h_1 \phi(z) \\
b_j & \rightarrow h_1 b_j \\
\nonumber
u & \rightarrow a_{0(cyl)}^- - 1/2 = a_{0(pl)}^- - 1
\end{align}
in (\ref{h1fromfermion}) to arrive at (\ref{h1expansionns}) with an extra overall factor modifying the operator for $q \neq 0$. For the $\mathcal{R}$-matrix the situation is even simpler because in this case we only need to change the sign of the $b_j$ modes to get an expression written in terms of $a_j^-$. Note that multiplication of the $\mathcal{R}$-matrix by a scalar function of the difference of the spectral parameters does not change the YBE.

To summarize this section, we found an alternative expression for the mixed $\mathcal{R}$-matrix (\ref{rmix}) such that
\begin{align}
\nonumber
\label{rfromfermion}
\tilde{\mathcal{R}}(u) & = \sum_{n=0}^\infty \frac{(-1)^{\frac{n(n+1)}{2}}}{n!} \prod_{j=1}^n \left[ \oint_0 \frac{dw_j}{2\pi i} \oint_{w_j} \frac{dz_j}{2\pi i} \kappa\left(u;z_j,w_j\right) \right] \times \\
& \times \frac{\prod_{j<k} (z_j-z_k) (w_j-w_k)}{\prod_{j,k} (z_j-w_k)} : e^{i\sum_j \phi_-(z_j) - i \sum_j \phi_-(w_j)}: \\
\nonumber
& = \frac{\Gamma(u+1)}{u^{-a_0^-} \Gamma(u+a_0^-+1)} \left[ \mathbbm{1} - \frac{\sum_{j_1>0} \mathcal{M}_{-j} \mathcal{M}_j}{u+a_0^-+1} + \frac{\sum_{j_1,j_2>0} \mathcal{M}_{-j_1,-j_2} \mathcal{M}_{j_1,j_2}}{2!(u+a_0^-+1)(u+a_0^-+2)} + \ldots \right]
\end{align}
Choosing the uncharged subsector $a_0^- = 0$ and using the replacement rule (\ref{fermrrepl}) reproduces the result (\ref{rmix}) found before. Note that the gamma-function prefactor has similar structure to the anomaly studied in \cite{Awata:2016bdm}. If we require the $\mathcal{R}$-matrix to preserve the highest weight state we miss this term, but it seems to naturally appear in the fermionic derivation of the $\mathcal{R}$-matrix.

Classically, there is a famous Szeg\"{o} formula comparing a determinantal formula analogous to (\ref{nsdeterminant}) to a contour integral of bosonized vertex operators like in (\ref{rfromfermion}) \cite{borodin2000fredholm}. We can thus interpret the transformation between these different generating functions of conserved quantities as a quantum version of the Szeg\"{o} formula.

\section{Arbesfeld-Schiffmann-Tsymbaliuk presentation}
\label{secastrelations}
In the previous section we studied the representation of the Yangian $\mathcal{R}$-matrix generators $\mathcal{H}$, $\mathcal{E}$ and $\mathcal{F}$ on single free boson Fock space. We found that $\mathcal{H}$ acts as (\ref{htaubosspectrum}). The generating function $\psi(u)$ of Tsymbaliuk presentation of Yangian acts as \cite{tsymbaliuk2017affine,Prochazka:2015deb}
\begin{equation}
\psi(u) \ket{\Lambda} = \frac{u-q + h_1 h_2 h_3 \psi_0}{u-q} \prod_{\Box \in \Lambda} \frac{(u-q-h_{\Box}+h_1)(u-q-h_{\Box}+h_2)(u-q-h_{\Box}+h_3)}{(u-q-h_{\Box}-h_1)(u-q-h_{\Box}-h_2)(u-q-h_{\Box}-h_3)} \ket{\Lambda}
\end{equation}
By comparing these two actions, we can identify
\begin{equation}
\label{psiHidentification}
\psi(u) = \frac{u-q + h_1 h_2 h_3 \psi_0}{u-q} \frac{\mathcal{H}^{(3)}(u+h_1)\mathcal{H}^{(3)}(u+h_2)}{\mathcal{H}^{(3)}(u)\mathcal{H}^{(3)}(u+h_1+h_2)}.
\end{equation}
The generating functions $\mathcal{H}^{(3)}(u)$ commute for all values of the spectral parameter $u$ so we can order them arbitrarily. Note that the right-hand side is written for $\tau = 3$ but the combination that appears is independent of $\tau$.

The generating functions $e(u)$ and $f(u)$ can be identified from (\ref{EHtoe1}) and \ref{EHtoe2}),
\begin{equation}
\label{EHtoe}
e(u) = h_\tau^{-1} \mathcal{E}^{(\tau)}(u-h_\tau) \left( \mathcal{H}^{(\tau)}(u-h_\tau) \right)^{-1} = h_\tau^{-1} \left( \mathcal{H}^{(\tau)}(u) \right)^{-1} \mathcal{E}^{(\tau)}(u).
\end{equation}
and (\ref{FHtof})
\begin{equation}
f(u) = -h_\tau^{-1} \left( \mathcal{H}^{(\tau)}(u-h_\tau) \right)^{-1} \mathcal{F}^{(\tau)}(u-h_\tau) = -h_\tau^{-1} \mathcal{F}^{(\tau)}(u) \left( \mathcal{H}^{(\tau)}(u) \right)^{-1}.
\end{equation}

We made this identifications based on the comparison of single boson representations, but the general form of the action is the same for all representations of MacMahon type obtained by taking a coproducts of single boson representations. Using the compatible coproduct guarantees that these identifications will hold also for any MacMahon representations. Note that the coproduct used in the $R$-matrix description of the algebra is the same one as the one used in \cite{Prochazka:2014gqa} as product of Miura operators. In \cite{Prochazka:2015deb} this coproduct was studied in the language of Tsymbaliuk's generators.

It is instructive to compare the Yangian algebra coming from Maulik-Okounkov $R$-matrix with Arbesfeld-Schiffmann-Tsymbaliuk presentation more explicitly. Assuming the large central charge expansion of the form
\begin{equation}
\psi(u) = 1 + \sigma_3 \frac{\psi_0}{u} + + \sigma_3 \frac{\psi_1}{u^2} + \ldots
\end{equation}
with $\psi_0$ and $\psi_1$ central and the analogous expansion
\begin{equation}
\mathcal{H}(u) = 1 + \frac{\mathcal{H}_1}{u} + \frac{\mathcal{H}_2}{u^2} + \ldots,
\end{equation}
the formula (\ref{psiHidentification}) allows us to find a triangular non-linear map between $\psi_j$ and $\mathcal{H}_j$ generators. The first few terms are
\begin{align}
\mathcal{H}_1 & = -\frac{h_3 \psi_2}{\psi_0} + \frac{h_3 \psi_1^2}{2\psi_0} \\
\nonumber
\mathcal{H}_2 & = -\frac{h_3 \psi_3}{6} + \frac{h_3^2(2h_1 h_2 \psi_0+3)\psi_2}{12} - \frac{h_3^2 \psi_1^2 \psi_2}{4\psi_0} + \frac{h_3^2\psi_2^2}{8} \\
& \; + \frac{h_3^2\psi_1^4}{8\psi_0^2} + \frac{h_3 \psi_1^3}{6\psi_0^2} - \frac{h_3^2(2h_1 h_2+\psi_0)\psi_1^2}{12\psi_0}
\end{align}
and where we identify the $\mathfrak{u}(1)$ charge $\psi_1$ with $\psi_0 q$. The first terms of the inverse transformation are
\begin{align}
\psi_1 & = q \psi_0 \\
\psi_2 & = q^2 \psi_0 - 2h_3^{-1} \mathcal{H}_1 \\
\psi_3 & = q^3 \psi_0 - (2h_1 h_2 \psi_0 + 3) \mathcal{H}_1 + 3h_3^{-1} \mathcal{H}_1^2 - 6h_3^{-1} \mathcal{H}_2.
\end{align}
The subalgebra generated by $\psi_j$ and by $\mathcal{H}_j$ is abelian.

\paragraph{$\left[\mathcal{H},\mathcal{E}\right]$ and $\left[\psi,e\right]$}
The next thing to compare are the $e_j$ and $\mathcal{E}_j$ generators. We use the left formula in (\ref{EHtoe}) and find
\begin{align}
\mathcal{E}_1 & = h_3 e_0 \\
\mathcal{E}_2 & = h_3 e_1 + h_3^2 e_0 \left(-\frac{\psi_2}{2} + \frac{\psi_1^2}{2\psi_0} - 1\right)
\end{align}
and the inverse
\begin{align}
e_0 & = h_3^{-1} \mathcal{E}_1 \\
e_1 & = h_3^{-1} \mathcal{E}_2 + h_3^{-1} \mathcal{E}_1 (h_3 - \mathcal{H}_1) \\
e_2 & = h_3^{-1} \mathcal{E}_3 + h_3^{-1} \mathcal{E}_2 (2h_3 - \mathcal{H}_1) + h_3^{-1} \mathcal{E}_1 (h_3^2 - 2h_3 \mathcal{H}_1 + \mathcal{H}_1^2 - \mathcal{H}_2).
\end{align}
First we compare the commutation relations of $\psi_j$ with $e_k$ and those of $\mathcal{H}_j$ and $\mathcal{E}_k$. The relations
\begin{align}
0 & = \left[ \psi_{j+3}, e_k \right] - 3 \left[ \psi_{j+2}, e_{k+1} \right] + 3\left[ \psi_{j+1}, e_{k+2} \right] - \left[ \psi_j, e_{k+3} \right] \\
& \quad + \sigma_2 \left[ \psi_{j+1}, e_k \right] - \sigma_2 \left[ \psi_j, e_{k+1} \right] - \sigma_3 \left\{ \psi_j, e_k \right\}
\end{align}
together with
\begin{align}
\left[ \psi_2, e_j \right] = 2 e_j
\end{align}
and the fact that $\psi_0$, $\psi_1$ are central allow us to calculate an arbitrary commutator $\left[\psi_j,e_k\right]$. The corresponding commutation relation in the Maulik-Okounkov algebra is (\ref{ybehe}) which in terms of modes reads
\begin{equation}
\left[ \mathcal{E}_{j+1}, \mathcal{H}_k \right] - \left[ \mathcal{E}_j, \mathcal{H}_{k+1} \right] = h_\tau (\mathcal{H}_j \mathcal{E}_k - \mathcal{H}_k \mathcal{E}_j) = h_\tau (\mathcal{E}_k \mathcal{H}_j - \mathcal{E}_j \mathcal{H}_k)
\end{equation}
This holds for any $j$ and $k$ with the convention that $\mathcal{E}_j = 0$ for $j \leq 0$, $\mathcal{H}_k = 0$ for $k<0$ and $\mathcal{H}_0 = 1$. In particular, the $j=0$ case is simply
\begin{equation}
\left[ \mathcal{H}_k, \mathcal{E}_1 \right] = -h_\tau \mathcal{E}_k
\end{equation}
and $k=0$ case is
\begin{equation}
\left[ \mathcal{H}_1, \mathcal{E}_j \right] = -h_\tau \mathcal{E}_j.
\end{equation}
For low values of $j$ and $k$ one can explicitly check that these equations are equivalent to those of Arbesfeld-Schiffmann-Tsymbaliuk for $\left[\psi_j,e_k\right]$ commutators (we checked these up to $j \leq 4$).

\paragraph{$\left[\mathcal{H},\mathcal{F}\right]$ and $\left[\psi,f\right]$}
The situation is analogous for the annihilation operators. The commutation relations of modes of $\mathcal{H}(u)$ and $\mathcal{E}(v)$ are
\begin{equation}
\left[ \mathcal{F}_{j+1}, \mathcal{H}_k \right] - \left[ \mathcal{F}_j, \mathcal{H}_{k+1} \right] = -h_\tau (\mathcal{H}_j \mathcal{F}_k - \mathcal{H}_k \mathcal{F}_j) = -h_\tau (\mathcal{F}_k \mathcal{H}_j - \mathcal{F}_j \mathcal{H}_k)
\end{equation}
The $j=0$ case gives
\begin{equation}
\left[\mathcal{H}_k,\mathcal{F}_1\right] = h_\tau \mathcal{F}_k
\end{equation}
and for $k=0$ we have
\begin{equation}
\left[\mathcal{H}_1,\mathcal{F}_j\right] = h_\tau \mathcal{F}_j
\end{equation}
Again, for lower mode numbers $j \leq 4$ we have checked that these relations are equivalent to the $\left[\psi_j,f_k\right]$ commutation relations of Arbesfeld-Schiffmann-Tsymbaliuk.

\paragraph{Other relations}
It would be nice to compare other relations that follow from Maulik-Okounkov $\mathcal{R}$-matrix to those of Arbesfeld-Schiffmann-Tsymbaliuk, but unfortunately this is not so easy to do in practice. One reason is that the $\mathcal{R}$-matrix in Fock representation has matrix elements labeled by pairs of partitions which is a very large set of generators. On the other hand, in Arbesfeld-Schiffmann-Tsymbaliuk presentation there are few generators (three generating functions $\psi(u)$, $e(u)$ and $f(u)$) but their relations are not strong enough to use an analogue of Poincar\'{e}-Birkhoff-Witt theorem directly. In fact, labeling the generators by non-negative spin and integer mode seems to be the optimal choice if we want to have a non-linear version of PBW theorem (i.e. a generalization of universal enveloping algebra with possible non-linear commutators). One possible generating set with this cardinality is to take the $U_{s,m}$ modes of $\mathcal{W}_{1+\infty}$. Another option is to consider the $h_1=h_2=h_3=0$ specialization of Arbesfeld-Schiffmann-Tsymbaliuk algebra which is equivalent to $w_{\infty}$ labeled again by spin and mode number. Since the generating functions $\psi(u), e(u)$ and $f(u)$ are summing the generators in the spin direction, it would be nice to complement Arbesfeld-Schiffmann-Tsymbaliuk generators $f(u), \psi(u), e(u)$ by an additional set of generating functions $e_j(u)$ and $f_j(u)$ with $j>1$. A simple possible candidate could be $\mathcal{E}_j(u)$ and $\mathcal{F}_j(u)$ fields as discussed in the appendix \ref{sechigheryangian}. While it seems that there is a homomorphism from the Maulik-Okounkov Yangian to Arbesfeld-Schiffmann-Tsymbaliuk Yangian, the existence of the map in the opposite direction is less clear, i.e. the Maulik-Okounkov Yangian is possibly larger than the Arbesfeld-Schiffmann-Tsymbaliuk Yangian. This issue surely deserves further study.

\section{Calogero-Moser-Sutherland models}
\label{seccalogero}
As an application of the $\mathcal{R}$-matrix formalism we will now study its connection to Calogero-Moser-Sutherland models. We will find simple vector representations of the Yangian (which can be useful to test the Yangian relations) and also a conformal field theoretic construction of quantum mechanical integrable models of Calogero type.

We will interpret the Miura operators (\ref{elementarymiura}) as being an $\mathcal{R}$-matrix in a mixed representation where one representation space is the bosonic Fock space while the other one is a space of functions of $z$ on which differential operators act. Instead of taking tensor product of $N$ Fock spaces with fixed $z$-space like we did until now, this time we consider $n$ $z$-spaces but a fixed Fock space, i.e. we consider the product
\begin{equation}
\mathcal{L}(z_1) \mathcal{L}(z_2) \ldots \mathcal{L}(z_n).
\end{equation}
By the general logic explained in chapter \ref{secrmatrix} the matrix elements of this operator should satisfy the relations of Yangian algebra derived from the Fock-Fock $\mathcal{R}$-matrix.

Let us first consider the case of cylinder. The mode expansion of the elementary Miura factor (\ref{elementarymiura}) is
\begin{equation}
-\frac{h_3}{h_1 h_2} \partial_z + J(z) = -\frac{h_3}{h_1 h_2} \partial_z + b_0 + \sum_{k \neq 0} b_k e^{-k z}.
\end{equation}
Here we expressed the parameter $\alpha_0$ in terms of $h_1, h_2$ and $h_3$ in such a way that $\mathcal{L}(z)$ is homogeneous with respect to rescaling of $h_j$. The zero mode $b_0$ is related to the spectral parameter by (\ref{spectparama0}) so up to an irrelevant constant shift of the spectral parameter we can put $u = -h_1 h_2 b_0$. Finally, for purposes of this section it is convenient to rescale the whole Miura factor such that its large $u$-expansion starts with the identity. The Miura factor that we will use in this section is thus
\begin{equation}
\mathcal{L}(z) = 1 + u^{-1} \left( h_3 \partial_z - h_1 h_2 \sum_{k \neq 0} b_k e^{-k z} \right).
\end{equation}
None of these rescalings or shifts of the spectral parameter affects the Yang-Baxter equation satisfied by $\mathcal{L}(z)$.

\paragraph{One insertion} The simplest situation is if we take just one Miura factor. The transfer matrix equals $\mathcal{L}(z)$. The generating function of Hamiltonians is
\begin{equation}
\mathcal{H}(u) = \bra{0} \mathcal{L}(z) \ket{0} = 1 + u^{-1} h_3 \partial_z
\end{equation}
while the elementary ladder operators are
\begin{align}
\mathcal{E}(u) = \bra{0} \mathcal{L}(z) b_{-1} \ket{0} = u^{-1} e^{-z} \\
\mathcal{F}(u) = \bra{0} b_1 \mathcal{L}(z) b_1 \ket{0} = u^{-1} e^z
\end{align}
We can now verify the Yangian commutation relations to check the consistency of our proposal. It is easy to see that the relations (\ref{ybehe2start}-\ref{ybehe2end}) are indeed satisfied. The Hamiltonian is diagonalized by the states of the form
\begin{equation}
\ket{k} = e^{kz}
\end{equation}
and the operators $\mathcal{E}$ and $\mathcal{F}$ act simply as ladder operators in this infinite dimensional space of periodic functions. Note that the representation that we obtain is has well-defined weight spaces with respect to $\mathcal{H}$ but it is not a highest weight representation. It is the vector representation discussed in \cite{tsymbaliuk2017affine}.

\paragraph{Two insertions} Picking two points on cylinder with coordinates $z_1$ and $z_2$ with $\Re z_2 > \Re z_1$, the transfer matrix is
\begin{equation}
\mathcal{T} = \mathcal{L}(z_2) \mathcal{L}(z_1)
\end{equation}
and the Hamiltonian
\begin{align}
\mathcal{H}(u) & = \bra{0} \mathcal{T} \ket{0} = 1 + \frac{h_3}{u} (\partial_{z_1} + \partial_{z_2}) + \frac{h_3^2}{u^2} \partial_{z_1} \partial_{z_2} -\frac{h_1 h_2}{u^2} \frac{e^{z_1+z_2}}{(e^{z_1}-e^{z_2})^2} \\
& = 1 + \frac{h_3}{u} (\partial_{z_1} + \partial_{z_2}) + \frac{h_3^2}{u^2} \partial_{z_1} \partial_{z_2} -\frac{h_1 h_2}{u^2} \frac{1}{4	\sinh^2\left(\frac{z_1-z_2}{2}\right)}.
\end{align}
We see that the coefficients of Taylor expansion in $b_0$ commute as follows from the Yang-Baxter equation. Interpreting this equation quantum mechanically, we see that we found a variant of P\"oschl-Teller potential for the relative motion of two particles. The term of order $u^{-1}$ is the total momentum while the $u^{-2}$ term is after addition of the square of the momentum equal to the energy. We can similarly find the ladder operators:
\begin{align}
\mathcal{E}(u) & = \bra{0} \mathcal{T} b_{-1} \ket{0} = \frac{1}{u} (e^{-z_1} + e^{-z_2}) + \frac{h_3}{u^2} (e^{-z_1} \partial_{z_2} + e^{-z_2} \partial_{z_1}) \\
\mathcal{F}(u) & = \bra{0} b_1 \mathcal{T} \ket{0} = \frac{1}{u} (e^{z_1} + e^{z_2}) + \frac{h_3}{u^2} (e^{z_1} \partial_{z_2} + e^{z_2} \partial_{z_1}).
\end{align}
Analogously to $n=1$ case, we don't expect to find a highest weight representation. But we can look at a state of vanishing total momentum (i.e. a wave-function depending only on $z_1-z_2$) with minimum energy. There are two such states,
\begin{equation}
\varphi_1(z_1,z_2) = \left[\sinh^2 \left( \frac{z_1-z_2}{2}\right) \right]^{-\frac{h_1}{h_3}} \quad \textrm{and} \quad \varphi_2(z_1,z_2) = \left[\sinh^2 \left( \frac{z_1-z_2}{2}\right) \right]^{-\frac{h_2}{h_3}}
\end{equation}
with $\mathcal{H}(u)$ eigenvalues
\begin{equation}
\frac{\left(u-\frac{h_1}{2}\right)\left(u+\frac{h_1}{2}\right)}{u^2} \quad\quad \textrm{and} \quad\quad \frac{\left(u-\frac{h_2}{2}\right)\left(u+\frac{h_2}{2}\right)}{u^2}.
\end{equation}
We may obtain other states by acting with ladder operators. For example, the action of $\mathcal{E}(u)$ on $\varphi_1(z_1,z_2)$ gives us wave function
\begin{equation}
(e^{-z_1} + e^{-z_2}) \varphi_1(z_1,z_2)
\end{equation}
which is an eigenstate of $\mathcal{H}(u)$ with eigenvalue
\begin{equation}
\frac{\left(u-\frac{h_1}{2}\right)\left(u+\frac{h_1}{2}-h_3\right)}{u^2}
\end{equation}
while the action of $\mathcal{F}(u)$ on $\varphi_1(z_1,z_2)$ gives
\begin{equation}
(e^{z_1} + e^{z_2}) \varphi_1(z_1,z_2)
\end{equation}
with $\mathcal{H}(u)$ eigenvalue
\begin{equation}
\frac{\left(u-\frac{h_1}{2}+h_3\right)\left(u+\frac{h_1}{2}\right)}{u^2}.
\end{equation}
Action of $\mathcal{E}(u)$ followed by the action of $\mathcal{F}(u)$ on $\varphi_1(z_1,z_2)$ results in a linear combination of the state $\varphi_1(z_1,z_2)$ and another state
\begin{equation}
\cosh(z_1-z_2) \varphi_1(z_1,z_2)
\end{equation}
with $\mathcal{H}(u)$ eigenvalue
\begin{equation}
\frac{\left(u-\frac{h_1}{2}+h_3\right)\left(u+\frac{h_1}{2}-h_3\right)}{u^2}.
\end{equation}
The simplest way how to decouple the states produced by the action of ladder operator into $\mathcal{H}(u)$ eigenstates is to use the ladder operators (\ref{EHtoe}) and (\ref{FHtof}) which have the property that the residues at their simple poles in $u$ are $\mathcal{H}$-eigenstates. This is the analogue of Bethe states for our algebra.

\paragraph{Multiple insertions} We can proceed similarly with any number of insertions of $\mathcal{L}(z)$. The ladder operators $\mathcal{E}(u)$ and $\mathcal{F}(u)$ as well as the Hamiltonian $\mathcal{H}(u)$ can be evaluated straightforwardly using the Wick theorem: the Hamiltonian takes the form
\begin{equation}
\label{calogeroH}
\mathcal{H}(u) = \sum_{\textrm{pairings}} \prod_{l \, \textrm{unpaired}} \left(1 + u^{-1} h_3 \partial_{z_l} \right) \prod_{\textrm{pairs} \, j<k} \left(-\frac{h_1 h_2}{u^2} \frac{1}{4	\sinh^2\left(\frac{z_j-z_k}{2}\right)} \right).
\end{equation}
The sum is over all possible partial pairings. The first product runs over all terms that are not paired while the second product is the contraction corresponding to contracted pairs. Incidentally, this kind of generating function of commuting higher Hamiltonians of $n$-particle Calogero-Sutherland model was discussed recently in \cite{Polychronakos:2018gfz}. The leading terms in the large $u$ expansion of $\mathcal{H}(u)$ are
\begin{align}
\mathcal{H}(u) & = 1 + \frac{h_3}{u} \sum_j \partial_{z_j} + \frac{1}{u^2} \sum_{j<k} \left[ h_3^2 \partial_{z_j} \partial_{z_k} -h_1 h_2 \frac{1}{4\sinh^2 \left(\frac{z_j-z_k}{2}\right)} \right] + \\
& + \frac{1}{u^3} \sum_{j<k<l} \left[ h_3^3 \partial_{z_j} \partial_{z_k} \partial_{z_l} - h_1 h_2 h_3 \frac{\partial_{z_l}}{4\sinh^2\left(\frac{z_j-z_k}{2}\right)} + cycl. \right] + \cdots
\end{align}
where in the last term we have three cyclic permutations of $(z_j,z_k,z_l)$.

One way of writing the ladder operators is to notice that in the case of $\mathcal{E}(z)$ the in-state $b_{-1} \ket{0}$ must be contracted with one of the $\mathcal{L}(z)$ operators after which the evaluation reduces to that of $\mathcal{H}(z)$ with one less variable. This means that we have
\begin{align}
\mathcal{E}(z) & = u^{-1} \sum_{j=1}^n e^{-z_j} \mathcal{H}_{\hat{j}}(z) \\
\mathcal{F}(z) & = u^{-1} \sum_{j=1}^n e^{z_j} \mathcal{H}_{\hat{j}}(z)
\end{align}
where we used the notation $\mathcal{H}_{\hat{j}}(u)$ for (\ref{calogeroH}) where the $j$-th coordinate is left out.

\paragraph{Plane} Instead of finding differential operators acting on the cylinder, we can transform to the plane via (\ref{cylindertoplane}). The differential operators for one insertion are
\begin{align}
\mathcal{H}(u) & = 1 + u^{-1} h_3 \tilde{z} \partial_{\tilde{z}} \\
\mathcal{E}(u) & = u^{-1} \tilde{z}^{-1} \\
\mathcal{F}(u) & = u^{-1} \tilde{z},
\end{align}
for two insertions
\begin{align}
\mathcal{H}(u) & = 1 + u^{-1} h_3 (\tilde{z}_1 \partial_{\tilde{z}_1} + \tilde{z}_2 \partial_{\tilde{z}_2}) + u^{-2} \left(h_3^2 \tilde{z}_1 \tilde{z}_2 \partial_{\tilde{z}_1} \partial_{\tilde{z}_2} - h_1 h_2 \frac{\tilde{z}_1 \tilde{z}_2}{(\tilde{z}_1-\tilde{z}_2)^2} \right) \\
\mathcal{E}(u) & = u^{-1} (\tilde{z}_1^{-1} + \tilde{z}_2^{-1}) + u^{-2} h_3 (\tilde{z}_1^{-1} \tilde{z}_2 \partial_{\tilde{z}_2} + \tilde{z}_2^{-1} \tilde{z}_1 \partial_{\tilde{z}_1}) \\
\mathcal{F}(u) & = u^{-1} (\tilde{z}_1 + \tilde{z}_2) + u^{-2} h_3 (\tilde{z}_1 \tilde{z}_2 \partial_{\tilde{z}_2} + \tilde{z}_2 \tilde{z}_1 \partial_{\tilde{z}_1})
\end{align}
etc. Returning to the case of one insertion, we can use the equations (\ref{EHtoe2}) and (\ref{FHtof}) to find the generating functions of Tsymbaliuk
\begin{align}
e(u) & = h_3^{-1} u^{-1} (1 + u^{-1} h_3 E)^{-1} \tilde{z}^{-1} \\
f(u) & = - h_3^{-1} u^{-1} \tilde{z} (1 + u^{-1} h_3 E)^{-1}
\end{align}
with Euler operator $E \equiv \tilde{z} \partial_{\tilde{z}}$. In terms of (spin) modes
\begin{align}
e_j & = (-1)^j h_3^{j-1} E^j \tilde{z}^{-1} \\
f_j & = (-1)^{j-1} h_3^{j-1} \tilde{z} E^j.
\end{align}
These allow us to determine the $\psi_j$ generators
\begin{align}
\psi_0 & = 0 \\
\psi_1 & = h_3^{-1} \\
\psi_2 & = 1 - 2E \\
\psi_3 & = h_3 (1-3E+3E^2) \\
\psi_4 & = h_3^2 (1-4E+6E^2-4E^3)
\end{align}
and in general
\begin{equation}
\psi_j = h_3^{j-2} \sum_{k=0}^{j-1} (-1)^k {j \choose k} E^k = h_3^{j-2} ((1-E)^j-(-E)^j).
\end{equation}
The corresponding generating function of $\psi$-charges is
\begin{equation}
\psi(u) = \frac{(u+h_3 E+h_1)(u+h_3 E+h_2)}{(u+h_3 E)(u+h_3 E-h_3)}.
\end{equation}
In contrast to representations of the affine Yangian on partitions and plane partitions where $\psi_0 \neq 0$, here $\psi_0 = 0$. This representation of the algebra is the vector representation of \cite{tsymbaliuk2017affine} and is the representation that acts on orbital degrees of freedom in Calogero models.

The transition from vector-like representations which have fixed number of Bethe roots to Fock representations physically corresponds to the transition to the second quantized picture (or grand-canonical ensemble) with variable number of particles (Bethe roots). The Fock oscillators can be identified with collective coordinates of the particles.

\paragraph{Torus}
Studying the $n$-point functions of the elementary Miura factor on the plane and cylinder we found the rational and trigonometric Calogero-Moser-Sutherland models. It is interesting to observe that if we replace in (\ref{calogeroH}) the two-point function
\begin{equation}
\frac{1}{4\sinh^2\left(\frac{z_j-z_k}{2}\right)}
\end{equation}
by its generalization on a torus which is the Weierstrass function, we find a generating function of quantum mechanical elliptic Calogero model, i.e.
\begin{equation}
\mathcal{H}(u) = \sum_{\textrm{pairings}} \prod_{l \, \textrm{unpaired}} \left(1 + u^{-1} h_3 \partial_{z_l} \right) \prod_{\textrm{pairs} \, j<k} \left(-\frac{h_1 h_2}{u^2} \wp_{\tau}\left(z_j-z_k\right) \right).
\end{equation}
The commutativity of these quantum Hamiltonians is not immediately guaranteed by the Yang-Baxter equation (because in the Hamiltonian formalism we are using the handle-insertion operator whose Yang-Baxter equation we have not verified), but still they commute as can be verified on first few Hamiltonians using the identities satisfied by elliptic functions. The commutativity of quadratic and cubic Hamiltonian requires for example
\begin{equation}
\det\begin{pmatrix} \wp(z_1-z_2) & \wp(z_2-z_3) & \wp(z_3-z_1) \\ \wp^\prime(z_1-z_2) & \wp^\prime(z_2-z_3) & \wp^\prime(z_3-z_1) \\ 1 & 1 & 1 \end{pmatrix}
\end{equation}
which is the well-known addition identity for Weierstrass elliptic function.

\paragraph{Summary}
In this section we interpreted the Miura transformation as a transfer matrix in the mixed representation where one of the vector spaces is the Fock space while the other one is a space of functions of worldsheet position $z$ with action of differential operators. The Yangian generators that we found were geometrically given by CFT correlation functions on a sphere or a cylinder with insertions of elementary Miura factors. The choice of in and out state specified the Yangian generator. The Yangian algebra itself was insensitive to positions of insertions of Miura factors or their number. The Yangian algebra is encoding the Ward identities for this class of correlators. There are few obvious generalizations that one can consider: instead of $N=1$ Fock space and Miura operator one could consider the higher rank case. The whole discussion should be entirely analogous. Another possible generalization would be to consider more complicated insertions. Interpreting the differential operators representing the Yangian as differential operators acting on moduli space of punctured spheres, one could also ask what happens if the genus is higher than zero.

\section{Outlook}
There are various possible directions in which the results of this article could be extended. We found two formulas for the $\mathcal{R}$-matrix of the mixed type, but if the Fock spaces are of the same type, the only explicit expression known so far is the fermionic formula derived by Smirnov \cite{Smirnov:2013hh}. Parts of that formula are quite reminiscent of the mixed $\mathcal{R}$-matrix at $\rho=1$ discussed here so perhaps the $\rho \neq 1$ bosonic $\mathcal{R}$-matrix is not out of the reach.

We also made only first steps in exploration of the structure of Maulik-Okounkov Yangian. We found a map from the Maulik-Okounkov Yangian to Arbesfeld-Schiffmann-Tsymbaliuk algebra, but it is not clear what is its kernel. It doesn't seem to be trivial, so in the optimistic case it could be central \footnote{Thanks to O. Schiffmann for discussing this issue.}. Focusing on simpler vector representations of the Yangian could possibly further simplify the problem of studying the $\mathcal{R}$-matrix.

\paragraph{Orthosymplectic case}
The $\mathcal{R}$-matrix considered here is associated to $\mathcal{W}$-algebras of Dynkin type $A_{N-1}$. But there are also orthosymplectic $\mathcal{W}$-algebras which are quotients of even spin subalgebra of $\mathcal{W}_{1+\infty}$. On the level of classical differential operators one can get the orthosymplectic spin chain by adding a reflection operator at one of the ends of the spin chain \cite{drinfeld1985lie}. Better understanding how this works in the quantum case could lead to an analogue of Arbesfeld-Schiffmann-Tsymbaliuk description of even spin $\mathcal{W}_{\infty ev}$.

\paragraph{Calogero models}
We found in the last section that that the correlation functions of rank $1$ Miura factors on a cylinder produce Hamiltonians and ladder operators of the trigonometric Calogero-Sutherland model. The generalization to higher ranks is straightforward, but it would be interesting to see if there is also a similar picture that would apply to higher genus Riemann surfaces or to different types of insertions. The Hitchin systems provide a large class of classical integrable models labeled by (punctured) Riemann surfaces together with a gauge group so it would be interesting to see which of those one can reproduce using the Maulik-Okounkov $\mathcal{R}$-matrix approach. In the conformal field theory we have Knizhink-Zamolodchikov-Bernard equations which take a form of differential equations on Riemann surfaces so one should see how these are related.

\paragraph{Classical integrability and KP hierarchy}
There is a well-developed theory of hierarchies of classical integrable partial differential equations \cite{miwa2000solitons,Zabrodin:2018uwz}. The $\mathcal{W}_N$-algebras considered here are quantization of the first equations of $KdV_N$ hierarchies. Just like $\mathcal{W}_{\infty}$ interpolates between all $\mathcal{W}_N$ algebras, there is a KP hierarchy containing all the $KdV_N$ hierarchies. There is a way of encoding the hierarchy into Hirota equations for tau function which in its symmetric formulation \cite{Zabrodin:2018uwz} is reminiscent of the triality-invariant variables like those of Arbesfeld-Schiffmann-Tsymbaliuk. It would be nice to understand what is the quantum analogue of all of these classical constructions.

\paragraph{q-deformed version} Everything discussed here has an analogue in the context of quantum toroidal or Ding-Iohara-Miki algebras \cite{feigin2012quantum,Feigin:2013fga,Feigin:2015raa}. In particular, the $\mathcal{R}$-matrix in $q$-deformed setting was studied in detail in a series of papers \cite{Awata:2016mxc,Awata:2016bdm,Awata:2018svb} and in \cite{Fukuda:2017qki}. Since there is a Macdonald $q$-deformed version of Nazarov-Sklyanin opertors \cite{nazarov2014macdonald}, it could be interesting to see if it could lead to any additional insights in combination with the well-developed story of \cite{Awata:2016mxc,Awata:2016bdm,Awata:2018svb}. It would be also useful to translate the results on shuffle algebras studied in detail in the $q$-deformed case \cite{Negut:2013cz,Negut:2015jza,Negut:2016dxr} to the rational setting. In particular, the shuffle algebra could be the right way to describe more general matrix elements of the $\mathcal{R}$-matrix studied here.

\section*{Acknowledgement}
I would like to thank to Igor Bertan, Lorenz Eberhardt, Ond\v{r}ej Hul\'{i}k, Bra\v{n}o Jur\v{c}o, Sergey Lukyanov, Yutaka Matsuo, Luca Mattiello, Miroslav Rap\v{c}\'{a}k, Ivo Sachs, Martin Schnabl, Olivier Schiffmann, Alessandro Sfondrini, Andrey Smirnov, Alexander Tabler, Tung Tran, Oleksandr Tsymbaliuk and Rui-Dong Zhu for useful discussions. Special thanks to Aleksei Litvinov who also independently derived many results of this article.

This research was supported by the DFG Transregional Collaborative Research Centre TRR 33 and the DFG cluster of excellence Origin and Structure of the Universe.

\appendix

\section{Higher order expressions for $R$-matrix}
\label{apprmat}
The fourth order expression for $R$-matrix (in terms of $J_-$ current) is
\begin{align}
\nonumber
R^{(4)} = \, & r^{(4)} + \frac{1}{24} \left(r^{(1)}\right)^4 + \frac{1}{2} \left( r^{(1)} r^{(3)} + r^{(3)} r^{(1)} + r^{(2)} r^{(2)} \right) \\
& + \frac{1}{6} \left( r^{(1)} r^{(1)} r^{(2)} + r^{(1)} r^{(2)} r^{(1)} + r^{(2)} r^{(1)} r^{(1)} \right)
\end{align}
and
\begin{align}
\nonumber
r^{(4)} = & \frac{1}{4} \sum_{j_1,j_2,j_3,j_4>0} \left(a_{-j_1} a_{-j_2} a_{-j_3} a_{-j_4} a_{j_1+j_2+j_3+j_4} + a_{-j_1-j_2-j_3-j_4} a_{j_1} a_{j_2} a_{j_3} a_{j_4} \right) \\
\nonumber
& + \frac{1}{2} \sum_{j_1+j_2=k_1+k_2+k_3} \left( a_{-j_1} a_{-j_2} a_{k_1} a_{k_2} a_{k_3} + a_{-k_1} a_{-k_2} a_{-k_3} a_{j_1} a_{j_2} \right) \\
& - \frac{\rho}{8} \sum_{j_1,j_2>0} \left( a_{-j_1} a_{-j_2} a_{j_1+j_2} + a_{-j_1-j_2} a_{j_1} a_{j_2} \right) \\
\nonumber
& + \frac{\rho(\rho+1)}{8} \sum_{j_1,j_2>0} (j_1^2 + j_2^2 + j_1 j_2) \left( a_{-j_1} a_{-j_2} a_{j_1+j_2} + a_{-j_1-j_2} a_{j_1} a_{j_2} \right)
\end{align}
At fifth order, we have
\begin{align}
\nonumber
r^{(5)} = & -\frac{1}{5} \sum_{j_1,j_2,j_3,j_4,j_5>0} \left(a_{-j_1} a_{-j_2} a_{-j_3} a_{-j_4} a_{-j_5} a_{j_1+j_2+j_3+j_4+j_5} + a_{-j_1-j_2-j_3-j_4-j_5} a_{j_1} a_{j_2} a_{j_3} a_{j_4} a_{j_5} \right) \\
\nonumber
& -\frac{1}{2} \sum_{j_1+j_2+j_3+j_4=k_1+k_2} \left(a_{-j_1} a_{-j_2} a_{-j_3} a_{-j_4} a_{k_1} a_{k_2} + a_{-k_1} a_{-k_2} a_{j_1} a_{j_2} a_{j_3} a_{j_4} \right) \\
\nonumber
& -\frac{2}{3} \sum_{j_1+j_2+j_3=k_1+k_2+k_3} a_{-j_1} a_{-j_2} a_{-j_3} a_{k_1} a_{k_2} a_{k_3} \\
\nonumber
& + \frac{\rho}{6} \sum_{j_1,j_2,j_3>0} \left( a_{-j_1} a_{-j_2} a_{-j_3} a_{j_1+j_2+j_3} + a_{-j_1-j_2-j_3} a_{j_1} a_{j_2} a_{j_3} \right) \\
& - \frac{\rho(\rho+1)}{6} \sum_{j_1,j_2,j_3>0} (j_1^2+j_2^2+j_3^2+j_1 j_2+j_1 j_3+j_2 j_3) \times \\
\nonumber
& \times \left( a_{-j_1} a_{-j_2} a_{-j_3} a_{j_1+j_2+j_3} + a_{-j_1-j_2-j_3} a_{j_1} a_{j_2} a_{j_3} \right) \\
\nonumber
& + \frac{\rho}{4} \sum_{j_1+j_2=k_1+k_2} a_{-j_1} a_{-j_2} a_{k_1} a_{k_2} - \frac{\rho(\rho+1)}{8} \sum_{j_1+j_2=k_1+k_2} (j_1^2+j_2^2+k_1^2+k_2^2) a_{-j_1} a_{-j_2} a_{k_1} a_{k_2} \\
\nonumber
& - \frac{\rho^2(3\rho^2+9\rho+4)}{240} \sum_{j>0} j^4 a_{-j} a_j + \frac{\rho^2(\rho+1)}{24} \sum_{j>0} j^2 a_{-j} a_j - \frac{\rho^2(\rho+6)}{240} \sum_{j>0} a_{-j} a_j
\end{align}

\section{Fock representation of $\mathcal{W}_{1+\infty}$}
\label{appfock}
Since we are using here slightly different normalization of of Fock oscillators than the one used in \cite{Prochazka:2015deb}, let us summarize here the generators of $\mathcal{W}_{1+\infty}$ in terms of the oscillators. For concreteness we choose the representation associated to the third direction and choose the zero mode $b_0$ to act as zero (it can be reintroduced easily by a spectral shift).
\begin{align}
\nonumber
\psi_0 & = -\frac{1}{h_1 h_2} \\
\nonumber
\psi_1 & = 0 \\
\nonumber
\psi_2 & = -2h_1 h_2 \sum_{j>0} b_{-j} b_j \\
e_0 & = b_{-1} \\
\nonumber
f_0 & = -b_1 \\
\nonumber
e_1 & = -h_1 h_2 \sum_{j>0} b_{-j-1} b_j \\
\nonumber
f_1 & = h_1 h_2 \sum_{j>0} b_{-j} b_{j+1}
\end{align}
Box addition amplitudes in this representation
\begin{align}
\nonumber
e_0 J_1 & = -\frac{h_2}{h_1-h_2} J_2 + \frac{h_1}{h_1-h_2} J_{1,1} \\
\nonumber
e_0 J_2 & = -\frac{h_2}{2h_1-h_2} J_{3} + \frac{2h_1}{2h_1-h_2} J_{2,1} \\
\nonumber
e_0 J_{1,1} & = -\frac{2h_2}{h_1-2h_2} J_{2,1} + \frac{h_1}{h_1-2h_2} J_{1,1,1} \\
e_0 J_3 & = -\frac{h_2}{3h_1-h_2} J_4 + \frac{3h_1}{3h_1-h_2} J_{3,1} \\
\nonumber
e_0 J_{2,1} & = -\frac{h_2(h_1-2h_2)}{2(h_1-h_2)^2} J_{3,1} - \frac{h_1 h_2}{(h_1-h_2)^2} J_{2,2} + \frac{h_1(2h_1-h_2)}{2(h_1-h_2)^2} J_{2,1,1} \\
\nonumber
e_0 J_{1,1,1} & = -\frac{3h_2}{h_1-3h_2} J_{2,1,1} + \frac{h_1}{h_1-3h_2} J_{1,1,1,1}
\end{align}
Box annihilation amplitudes are instead
\begin{align}
\nonumber
f_0 J_1 & = \frac{1}{h_1 h_2} J_{\bullet} \\
\nonumber
f_0 J_2 & = \frac{2}{h_1 h_2} J_1 \\
\nonumber
f_0 J_{1,1} & = \frac{2}{h_1 h_2} J_1 \\
\nonumber
f_0 J_3 & = \frac{3}{h_1 h_2} J_2 \\
\nonumber
f_0 J_{2,1} & = \frac{h_1-2h_2}{h_1 h_2 (h_1-h_2)} J_2 + \frac{2h_1-h_2}{h_1 h_2(h_1-h_2)} J_{1,1} \\
f_0 J_{1,1,1} & = \frac{3}{h_1 h_2} J_{1,1} \\
\nonumber
f_0 J_4 & = \frac{4}{h_1 h_2} J_3 \\
\nonumber
f_0 J_{3,1} & = \frac{2(h_1-h_2)}{h_1 h_2(2h_1-h_2)} J_3 + \frac{6h_1-2h_2}{h_1 h_2(2h_1-h_2)} J_{2,1} \\
\nonumber
f_0 J_{2,2} & = \frac{4}{h_1 h_2} J_{2,1} \\
\nonumber
f_0 J_{2,1,1} & = \frac{2(h_1-3h_2)}{h_1 h_2(h_1-2h_2)} J_{2,1} + \frac{2h_1-2h_2}{h_1 h_2(h_1-2h_2)} J_{1,1,1} \\
\nonumber
f_0 J_{1,1,1,1} & = \frac{4}{h_1 h_2} J_{1,1,1}
\end{align}

\section{Higher relations of Yangian algebra}
\label{sechigheryangian}

\subsection{$\left[\mathcal{E},\mathcal{F}\right]$ relations}
For an illustration, let's discuss higher order relations of FZ algebra. Analogously to $\mathcal{H}, \mathcal{E}$ and $\mathcal{F}$, we can define
\begin{equation}
\mathcal{H}^{(\tau)}_{\Box} = \bra{0}_A a_{A,1} \mathcal{T}^{(\tau)}_{\Box,A} a_{A,-1} \ket{0}_A.
\end{equation}
The Yang-Baxter equation implies relations
\begin{multline}
(u_A-u_B+h_A) (u_A-u_B-h_B) \mathcal{H}^A \mathcal{H}^B_{\Box} = -h_B^2 \mathcal{H}^B \mathcal{H}^A_{\Box} + (u_A-u_B)(u_A-u_B+h_A-h_B) \mathcal{H}^B_{\Box} \mathcal{H}^A \\
- h_B(u_A-u_B) \mathcal{E}^B \mathcal{F}^A + h_B(u_A-u_B+h_A-h_B) \mathcal{F}^B \mathcal{E}^A
\end{multline}
and
\begin{multline}
(u_A-u_B+h_A) (u_A-u_B-h_B) \mathcal{H}^A_{\Box} \mathcal{H}^B = (u_A-u_B)(u_A-u_B+h_A-h_B) \mathcal{H}^B \mathcal{H}^A_{\Box} - h_A^2 \mathcal{H}^B_{\Box} \mathcal{H}^A \\
+ h_A(u_A-u_B) \mathcal{E}^B \mathcal{F}^A -h_A(u_A-u_B+h_A-h_B) \mathcal{F}^B \mathcal{E}^A.
\end{multline}
Permuting labels $A \leftrightarrow B$ we find additional two relations
\begin{multline}
(u_A-u_B+h_A) (u_A-u_B-h_B) \mathcal{H}^B \mathcal{H}^A_{\Box} = (u_A-u_B)(u_A-u_B+h_A-h_B) \mathcal{H}^A_{\Box} \mathcal{H}^B -h_A^2 \mathcal{H}^A \mathcal{H}^B_{\Box} \\
+h_A(u_A-u_B) \mathcal{E}^A \mathcal{F}^B -h_A(u_A-u_B+h_A-h_B) \mathcal{F}^A \mathcal{E}^B
\end{multline}
and
\begin{multline}
(u_A-u_B+h_A) (u_A-u_B-h_B) \mathcal{H}^B_{\Box} \mathcal{H}^A = (u_A-u_B)(u_A-u_B+h_A-h_B) \mathcal{H}^A \mathcal{H}^B_{\Box} - h_B^2 \mathcal{H}^A_{\Box} \mathcal{H}^B \\
-h_B(u_A-u_B) \mathcal{E}^A \mathcal{F}^B +h_B(u_A-u_B+h_A-h_B) \mathcal{F}^A \mathcal{E}^B.
\end{multline}
We also have relations
\begin{multline}
(u_A-u_B+h_A) (u_A-u_B-h_B) \mathcal{E}^A \mathcal{F}^B = -h_B(u_A-u_B+h_A-h_B) \mathcal{H}^B \mathcal{H}^A_{\Box} + h_A(u_A-u_B+h_A-h_B) \mathcal{H}^B_{\Box} \mathcal{H}^A \\
-h_A h_B \mathcal{E}^B \mathcal{F}^A + (u_A-u_B+h_A-h_B)^2 \mathcal{F}^B \mathcal{E}^A
\end{multline}
and
\begin{multline}
(u_A-u_B+h_A) (u_A-u_B-h_B) \mathcal{F}^A \mathcal{E}^B = h_B(u_A-u_B) \mathcal{H}^B \mathcal{H}^A_{\Box} - h_A(u_A-u_B) \mathcal{H}^B_{\Box} \mathcal{H}^A \\
+(u_A-u_B)^2 \mathcal{E}^B \mathcal{F}^A -h_A h_B \mathcal{F}^B \mathcal{E}^A.
\end{multline}
as well as their $A \leftrightarrow B$ permutations
\begin{multline}
(u_A-u_B+h_A) (u_A-u_B-h_B) \mathcal{E}^B \mathcal{F}^A = h_A(u_A-u_B+h_A-h_B) \mathcal{H}^A \mathcal{H}^B_{\Box} - h_B(u_A-u_B+h_A-h_B) \mathcal{H}^A_{\Box} \mathcal{H}^B \\
-h_A h_B \mathcal{E}^A \mathcal{F}^B + (u_A-u_B+h_A-h_B)^2 \mathcal{F}^A \mathcal{E}^B
\end{multline}
and
\begin{multline}
(u_A-u_B+h_A) (u_A-u_B-h_B) \mathcal{F}^B \mathcal{E}^A = -h_A(u_A-u_B) \mathcal{H}^A \mathcal{H}^B_{\Box} + h_B(u_A-u_B) \mathcal{H}^A_{\Box} \mathcal{H}^B \\
+(u_A-u_B)^2 \mathcal{E}^A \mathcal{F}^B -h_A h_B \mathcal{F}^A \mathcal{E}^B.
\end{multline}
Out of these $8$ relations for $8$ unknowns, only $4$ are independent. We can eliminate from these equations $\mathcal{H}_{\Box}$ (and thus also $\mathcal{H}$) and find
\begin{equation}
\label{EFrelation}
(u_A-u_B) (\mathcal{E}^A \mathcal{F}^B - \mathcal{E}^B \mathcal{F}^A) = (u_A-u_B+h_A-h_B) (\mathcal{F}^B \mathcal{E}^A - \mathcal{F}^A \mathcal{E}^B)
\end{equation}
Note that this equation is trivially satisfied for $A=B$, i.e. for auxiliary spaces of the same type. Analogously we can derive relations
\begin{align}
h_A \left[\mathcal{H}^A,\mathcal{H}^B_{\Box}\right] & = h_B \left[\mathcal{H}^B, \mathcal{H}^A_{\Box}\right] \\
(u_A-u_B) \mathcal{E}^B \mathcal{F}^A - (u_A-u_B+h_A-h_B) \mathcal{F}^A \mathcal{E}^B & = h_A \mathcal{H}^B_{\Box} \mathcal{H}^A - h_B \mathcal{H}^A_{\Box} \mathcal{H}^B \\
h_B \left[ \mathcal{E}^A, \mathcal{F}^B \right] + h_B \left[\mathcal{F}^A, \mathcal{E}^B \right] & = (h_A-h_B) \left[ \mathcal{H}^A, \mathcal{H}^B_{\Box} \right] \\
(u_A-u_B+h_A-h_B) \left[\mathcal{H}^A, \mathcal{H}^B_{\Box} \right] & = h_B(\mathcal{E}^A \mathcal{F}^B - \mathcal{E}^B \mathcal{F}^A)
\end{align}

\subsection{$\left[\mathcal{E},\mathcal{E}\right]$ relations}
Consider now two additional operators by taking the matrix elements
\begin{equation}
\mathcal{E}^{(\tau)}_{2} = \bra{0}_A \mathcal{T}^{(\tau)}_A a_{A,-2} \ket{0}_A
\end{equation}
and
\begin{equation}
\mathcal{E}^{(\tau)}_{1,1} = \bra{0}_A \mathcal{T}^{(\tau)}_A a_{A,-1}^2 \ket{0}_A.
\end{equation}
To derive the commutation relations between these and $\mathcal{H}^{(\tau)}$, we need to know the matrix elements of $\mathcal{R}_{AB}$ at level $2$.	It is straightforward to derive these from (\ref{ractlevel2}) but since the result is quite complicated, let us specialize for simplicity to $\tau_A = \tau_B = \tau$. In this case we have
\begin{align}
\nonumber
\Delta \mathcal{R}_{AB} a_{A,-2} \ket{0} & = (u_A-u_B)(h_\tau(u_A-u_B+h_\tau)^2+\sigma_3) a_{A,-2} -h_\tau(u_A-u_B)\sigma_3 a_{A,-1}^2 \\
\nonumber
& + h_\tau(2h_\tau(u_A-u_B+h_\tau)^2+\sigma_3) a_{B,-2} - h_\tau(u_A-u_B)\sigma_3 a_{B,-1}^2 \\
& + 2h_\tau(u_A-u_B)\sigma_3 a_{A,-1}a_{B,-1} \\
\nonumber
\Delta \mathcal{R}_{AB} a_{B,-2} \ket{0} & = h_\tau(2h_\tau(u_A-u_B+h_\tau)^2+\sigma_3) a_{A,-2} + h_\tau(u_A-u_B)\sigma_3 a_{A,-1}^2 \\
\nonumber
& + (u_A-u_B)(h_\tau(u_A-u_B+h_\tau)^2+\sigma_3) a_{B,-2} + h_\tau(u_A-u_B)\sigma_3 a_{B,-1}^2 \\
& - 2h_\tau(u_A-u_B)\sigma_3 a_{A,-1}a_{B,-1} \\
\nonumber
\Delta \mathcal{R}_{AB} a_{A,-1}^2 \ket{0} & = h_\tau^2(u_A-u_B) a_{A,-2} + (u_A-u_B)(h_\tau(u_A-u_B+h_\tau)^2+\sigma_3-h_\tau^3) a_{A,-1}^2 \\
\nonumber
& - h_\tau^2(u_A-u_B) a_{B,-2} + h_\tau(h_\tau^2(u_A-u_B+2h_\tau)+\sigma_3) a_{B,-1}^2 \\
& + 2h_\tau^2(u_A-u_B)(u_A-u_B+2h_\tau) a_{A,-1} a_{B,-1} \\
\nonumber
\Delta \mathcal{R}_{AB} a_{B,-1}^2 \ket{0} & = h_\tau^2(u_A-u_B)a_{A,-2} + h_\tau(h_\tau^2(u_A-u_B+2h_\tau)+\sigma_3) a_{A,-1}^2 \\
\nonumber
& - h_\tau^2(u_A-u_B) a_{B,-2} + (u_A-u_B)(h_\tau(u_A-u_B+h_\tau)^2+\sigma_3-h_\tau^3) a_{B,-1}^2 \\
& + 2h_\tau^2(u_A-u_B)(u_A-u_B+2h_\tau) a_{A,-1} a_{B,-1} \\
\nonumber
\Delta \mathcal{R}_{AB} a_{A,-1} a_{B,-1} \ket{0} & = -h_\tau^2(u_A-u_B) a_{A,-2} + h_\tau^2(u_A-u_B)(u_A-u_B+2h_\tau) a_{A,-1}^2 \\
& + h_\tau^2(u_A-u_B) a_{B,-2} + h_\tau^2(u_A-u_B) (u_A-u_B+2h_\tau) a_{B,-1}^2 \\
\nonumber
& + (\sigma_3(u_A-u_B+h_\tau)+h_\tau(u_A-u_B+2h_\tau)((u_A-u_B)^2+h_\tau^2)) a_{A,-1} a_{B,-1}
\end{align}
where
\begin{equation}
\Delta = (u_A-u_B+h_\tau)\left[h_\tau(u_A-u_B+h_\tau)(u_A-u_B+2h_\tau)+\sigma_3\right]
\end{equation}
The level 2 relations are then
\begin{align}
\nonumber
\Delta \mathcal{E}_2(u) \mathcal{H}(v) & = (u-v)(h_\tau(u-v+h_\tau)^2+\sigma_3) \mathcal{H}(v) \mathcal{E}_2(u) -h_\tau(u-v)\sigma_3 \mathcal{H}(v) \mathcal{E}_{1,1}(u) \\
\nonumber
& + h_\tau(2h_\tau(u-v+h_\tau)^2+\sigma_3) \mathcal{E}_2(v) \mathcal{H}(u) - h_\tau(u-v)\sigma_3 \mathcal{E}_{1,1}(v) \mathcal{H}(u) \\
& + 2h_\tau(u-v)\sigma_3 \mathcal{E}(v) \mathcal{E}(u) \\
\nonumber
\Delta \mathcal{H}(u) \mathcal{E}_2(v) & = h_\tau(2h_\tau(u-v+h_\tau)^2+\sigma_3) \mathcal{H}(v) \mathcal{E}_2(u) + h_\tau(u-v)\sigma_3 \mathcal{H}(v) \mathcal{E}_{1,1}(u) \\
\nonumber
& + (u-v)(h_\tau(u-v+h_\tau)^2+\sigma_3) \mathcal{E}_2(v) \mathcal{H}(u) + h_\tau(u-v)\sigma_3 \mathcal{E}_{1,1}(v) \mathcal{H}(u) \\
& - 2h_\tau(u-v)\sigma_3 \mathcal{E}(v) \mathcal{E}(u) \\
\nonumber
\Delta \mathcal{E}_{1,1}(u) \mathcal{H}(v) & = h_\tau^2(u-v) \mathcal{H}(v) \mathcal{E}_2(u) + (u-v)(h_\tau(u-v+h_\tau)^2+\sigma_3-h_\tau^3) \mathcal{H}(v) \mathcal{E}_{1,1}(u) \\
\nonumber
& - h_\tau^2(u-v) \mathcal{E}_2(v) \mathcal{H}(u) + h_\tau(h_\tau^2(u-v+2h_\tau)+\sigma_3) \mathcal{E}_{1,1}(v) \mathcal{H}(u) \\
& + 2h_\tau^2(u-v)(u-v+2h_\tau) \mathcal{E}(v) \mathcal{E}(u) \\
\nonumber
\Delta \mathcal{H}(u) \mathcal{E}_{1,1}(v) & = h_\tau^2(u-v) \mathcal{H}(v) \mathcal{E}_2(u) + h_\tau(h_\tau^2(u-v+2h_\tau)+\sigma_3) \mathcal{H}(v) \mathcal{E}_{1,1}(u) \\
\nonumber
& - h_\tau^2(u-v) \mathcal{E}_2(v) \mathcal{H}(u) + (u-v)(h_\tau(u-v+h_\tau)^2+\sigma_3-h_\tau^3) \mathcal{E}_{1,1}(v) \mathcal{H}(u) \\
& + 2h_\tau^2(u-v)(u-v+2h_\tau) \mathcal{E}(v) \mathcal{E}(u) \\
\nonumber
\Delta \mathcal{E}(u) \mathcal{E}(v) & = -h_\tau^2(u-v) \mathcal{H}(v) \mathcal{E}_2(u) + h_\tau^2(u-v)(u-v+2h_\tau) \mathcal{H}(v) \mathcal{E}_{1,1}(u) \\
\nonumber
& + h_\tau^2(u-v) \mathcal{E}_2(v) \mathcal{H}(u) + h_\tau^2(u-v) (u-v+2h_\tau) \mathcal{E}_{1,1}(v) \mathcal{H}(u) \\
& + (\sigma_3(u-v+h_\tau)+h_\tau(u-v+2h_\tau)((u-v)^2+h_\tau^2)) \mathcal{E}(v) \mathcal{E}(u)
\end{align}
We can find another $5$ relations if we exchange $u \leftrightarrow v$. In total, we have 10 homogeneous linear equations for 10 unknowns, but three of the equations are dependent (the matrix of the linear system has rank $5$). We can eliminate some variables and find
\begin{equation}
\left[ \mathcal{H}(u), \mathcal{E}_2(v) \right] = \left[\mathcal{H}(v), \mathcal{E}_2(u) \right]
\end{equation}
(which is a consequence of commutativity of $\mathcal{R}$ with $J_+$),
\begin{equation}
(u-v) \left[ \mathcal{H}(u), \mathcal{E}_2(v)\right] +2h_3 \left( \mathcal{H}(u) \mathcal{E}_2(v) - \mathcal{H}(v) \mathcal{E}_2(u) \right) = h_1 h_2 \left[ \mathcal{E}(u), \mathcal{E}(v) \right]
\end{equation}
generalizing (\ref{ybehe}) or
\begin{multline}
(2h_3(u-v+h_3)+h_1 h_2)\mathcal{E}(u)\mathcal{E}(v) + (2h_3(u-v-h_3)-h_1 h_2)\mathcal{E}(v) \mathcal{E}(u) = \\
= 2(u-v)((u-v+h_3)\mathcal{H}(u)\mathcal{E}_{1,1}(v)-(u-v-h_3)\mathcal{E}_{1,1}(v)\mathcal{H}(u)) - (u-v) \left[\mathcal{H}(u),\mathcal{E}_2(v)\right].
\end{multline}

\bibliography{winfinstrm}

\end{document}